*Review*

# A Comprehensive Survey on the Internet of Things with the Industrial Marketplace

**Kazhan Othman Mohammed Salih** [1], **Tarik A. Rashid** [2,*], **Dalibor Radovanovic** [3] **and Nebojsa Bacanin** [3,*]

1   Project Management Department, College of Commerce, University of Sulaimani,
    Sulaimani 46001, Iraq; kazhan.mohammed@univsul.edu.iq
2   Computer Science and Engineering, School of Science and Engineering, University of Kurdistan Hewler,
    Erbil 44001, Iraq
3   Departman of Informatics and Computing, Faculty of Informatics and Computing, Singidunum University,
    Danijelova 32, 11000 Belgrade, Serbia; dradovanovic@singidunum.ac.rs
*   Correspondence: tarik.ahmed@ukh.edu.krd (T.A.R.); nbacanin@singidunum.ac.rs (N.B.)

**Abstract:** There is no doubt that new technology has become one of the crucial parts of most people's lives around the world. By and large, in this era, the Internet and the Internet of Things (IoT) have become the most indispensable parts of our lives. Recently, IoT technologies have been regarded as the most broadly used tools among other technologies. The tools and the facilities of IoT technologies within the marketplace are part of Industry 4.0. The marketplace is too regarded as a new area that can be used with IoT technologies. One of the main purposes of this paper is to highlight using IoT technologies in Industry 4.0, and the Industrial Internet of Things (IIoT) is another feature revised. This paper focuses on the value of the IoT in the industrial domain in general; it reviews the IoT and focuses on its benefits and drawbacks, and presents some of the IoT applications, such as in transportation and healthcare. In addition, the trends and facts that are related to the IoT technologies on the marketplace are reviewed. Finally, the role of IoT in telemedicine and healthcare and the benefits of IoT technologies for COVID-19 are presented as well.

## 1. Introduction

The human connection is one of the most important types of communication available on the Internet. The Internet is a global network that allows humans from all over the world to communicate with one another. A human-organized network of computers and smartphones is required to access the Internet. Personal computers and smartphones are examples of such devices. As a result, it has come to be known as "the Internet of People".

In a matter of a few years upcoming, the Internet will grow and evolve into the Internet of Things (IoT), shifting away from human-to-human communication and toward computers interacting with one another, which is also known as machine-to-machine (M2M) communication [1]. This new mode of communication will have a profound impact on the way people communicate and compute in their daily lives.

Kevin Ashton was the first who, in 1999, used the term Internet of Things [2]. Since 2008, the total number of things connected to the IoT has outnumbered the total number of people living on the planet. The IoT is expected to make it possible for everyone to access and provide a vast amount of information about equipment and locations in the future.

The IoT can be used to connect a variety of objects to a network, including artificial articles, plants, and animals. This can be counted as one of the fundamental differences between the IoT and the rest of the world. The IoT is still based on the Internet, but it is the most logical evolutionary and technological step forward for the Internet in recent history.

In this paper, we examine the origins and history of the IoT, and then we discuss IoT applications that have been developed, tested, and refined in the industrial marketplace. Our research motivation is spread across a broad range of areas within IoT technologies, rather than being restricted to a single specific zone.

The incorporation of the IoT and services into the business world heralds the beginning of the Fourth Industrial Revolution: Industry 4.0 [3]. The automotive industry is one of the fastest-growing IoT sectors [4–7]. Review studies on IoT for the automotive industry have also previously focused on specific issues like inter-vehicle networking [4,8], a comprehensive review of key technologies, and a succinct description of specific automotive IoT use cases [4,9].

While our survey focuses on some critical points in IoT technologies, other related topics are covered as well. According to our knowledge, no survey focusing on the IoT and COVID-19 has been conducted; therefore, we present the role of the IoT in medical science, and some of the benefits of the IoT in telemedicine in dealing with COVID-19 have been reviewed and investigated. Even though COVID-19 has been reviewed and investigated, there have been no papers published in this area as of yet. In addition to creating new challenges, the COVID-19 pandemic has also increased the pressure on healthcare systems around the world to adhere to stricter timelines for patient assessment, prescription, treatment, and health guidelines [10–13]. Moreover, the IoT is discussed in detail, including its advantages and disadvantages, along with Industry 4.0 and the roles of the IoT and IIoT in the marketplace.

The remainder of the paper will be divided into the following sections: Section 2 defines the IoT and describes its characteristics. Section 3 discusses the history of the IoT as well as current research in the field. Section 4 discusses several of the advantages and disadvantages of the IoT. Section 5 examines the architectural design of IoT technologies. Section 6 identifies and clarifies the trends, characteristics, and applications of the IoT. The IoT and pandemic control, which are relatively new topics these days, are linked in Section 7. Section 8 examines Industry 4.0 and the IIoT, as well as the intersection of the IoT and the industrial sector. Section 9 explains the IoT marketplace and how it is related to other marketplaces. Finally, in Section 10, the conclusion is presented, as well as recommendations for future work.

## 2. Internet of Things

There are many definitions of the IoT that have been made by many researchers before. Due to the convergence of multiple technologies, such as real-time analytics, machine learning, commodity sensors, and embedded systems, the definition of the IoT has evolved [14].

The IoT has been outlined as "Things that have identities and virtual personalities operating in smart spaces using intelligent interfaces to connect and communicate within the social, environmental, and user contexts" [1].

Through using IoT technologies, objects from the physical area can join each other and objects from the virtual area, hence facilitating anytime, anywhere connectivity for anything, not just for anyone [15,16]. The IoT has garnered noteworthy consideration from both the industry and academia [16]. It has caught the attention of industries, enterprises, governmental organizations, and the academic communities [17].

The key potency of the IoT concept is the strong effect it would have on numerous diverse sides of daily life and future users' behavior [18]. Another aspect of the revolution of the IoT is interconnecting humans at any time or from any place and enabling them to communicate with objects through communication networks. This is the underlying vision of the IoT [19]. Figure 1 illustrates the features of the IoT [1].

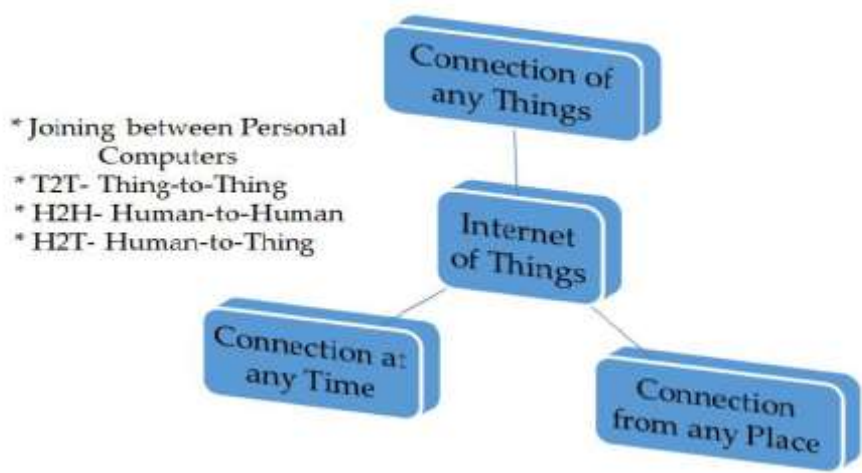

**Figure 1.** Features of the IoT.

As indicated by [20], the IoT, a rising worldwide Internet-based specialized structural design promoting the exchange of merchandise and ventures in worldwide flexible chain systems affects the security, protection, and confidentiality of the included partners [3,20].

In past years, industry fields needed to be monitored by a human, but since IoT technologies have been introduced, industry sectors have begun to be controlled, checked, and monitored by the IoT; human interference is not needed for these tasks and purposes. Within IoT technologies, the interconnection between humans is expanded to humans and things, or things and things [21,22], as shown in Figure 2, which determines various categories of the IoT.

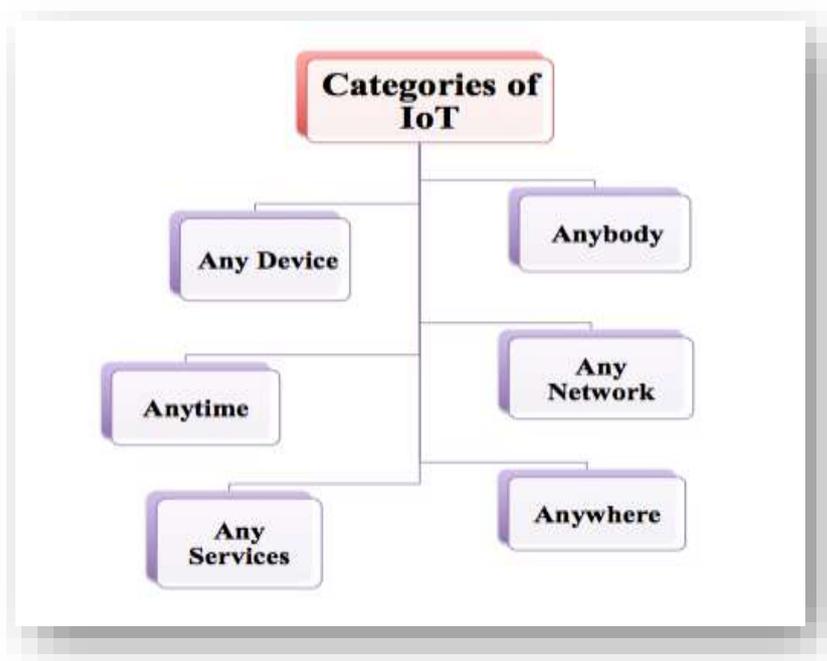

**Figure 2.** The IoT categories.

This means that communication may arise among smart devices rather than people. Recently, IoT technologies have begun to be used in various application domains, for instance, healthcare, systems of intelligent transportation and logistics, smart cities, the energy industry, etc. [23,24].

In IoT solutions, the requirements and technical capabilities of companies are being balanced constantly, and due to the IoT, as Figure 3 demonstrates, devices have been getting smarter and more connected in the last few decades [25]:

1. Connect

    Connectivity and processing efficiencies are built into equipment with IoT solutions.

2. Collect

    Sensors and storage capabilities are introduced as appliances are paired. Devices will also have a better overview of their surroundings and can gather and evaluate data for valuable information.

3. Compute

    Obtained data are identified, analyzed, assembled, and stocked in a big data solution.

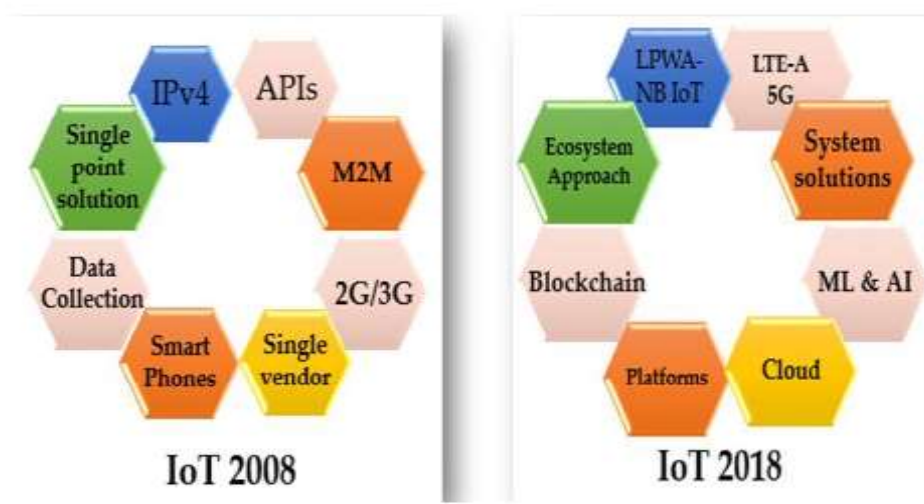

**Figure 3.** Due to the IoT, devices have been getting smarter and more connected in recent decades.

### 3. Background and Present Research on the IoT

Use of IoT technologies has increased rapidly within many industries. Using IoT technologies is accordingly required to improve the information for sharing within and across various industries [26]. Perceptions of IoT technologies have been pragmatic toward various applications ranging from home automation to industrial IoT, and to being able to communicate physical things from any place via a network [27]. It can be noted that the number of projects that relate to the use of the IoT in industrial areas has increased, for instance, agriculture, security surveillance, environmental monitoring, food processing, and many others. Moreover, published papers in the IoT areas have expanded considerably, spanning many different types of applications.

Many other technologies and strategies have, in addition, been used to support the IoT, such as Wi-Fi, ZigBee, service-oriented architecture (SOA), cloud computing (CC), radio-frequency identification (RFID), RFID CMOS, smartphones, social networks, and near field communication (NFC), as shown in Figure 4, which demonstrates the IoT interconnected with other technologies.

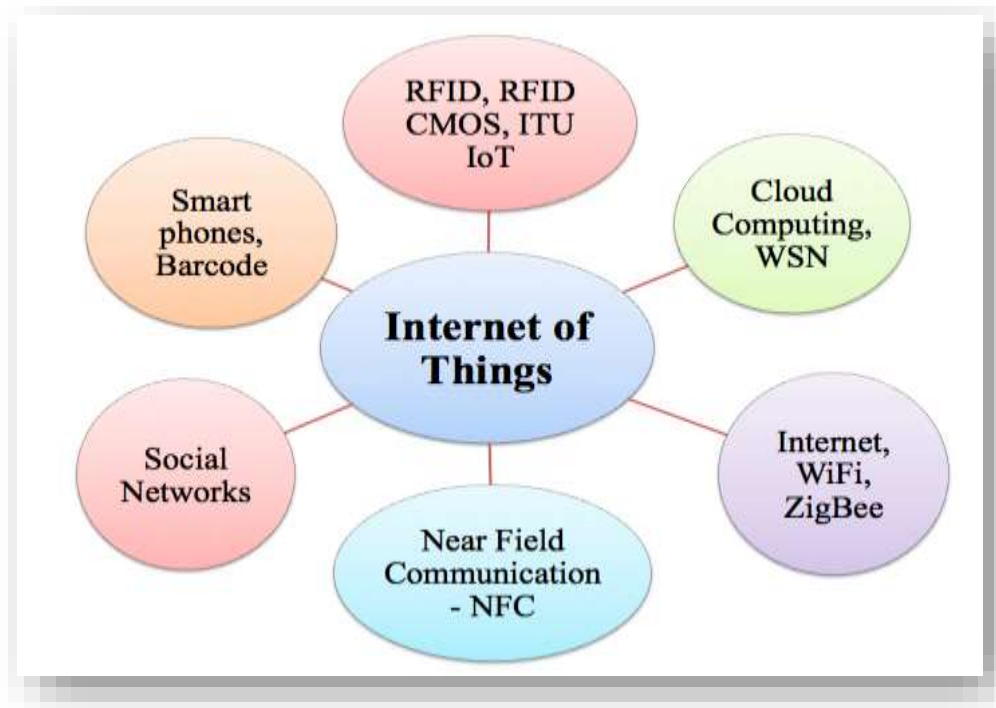

**Figure 4.** IoT interconnected with other technologies.

In the world of IoT technology, it is possible to identify six essential technologies that can be of great assistance to the manufacturing industry areas in the IoT. The following is a description of these:

1.  Radio-Frequency Identification (RFID)

    Primarily, the term "IoT" was introduced to indicate the particularly recognizable interoperable associated objects with radio-frequency identification (RFID) technology [2]. RFID makes the most of electromagnetic fields to pass data to automatically identify and track tags attached to objects [28,29]. RFID innovation is one of numerous IoT "empowering identification" technologies, and not essentially the major one [30]. It is one of the approaches used to link everyday items into networks [16,31]. The RFID system contains two main parts, which are:

    1.1. RFID tags: Enclosed items that consist of information regarding the items. RFID tags, which are sometimes called transponders (transmitters/responders), are attached to objects to count or identify them [32]. RFID tags, either passive or active, contain an antenna and microchip as well.

    1.2. Readers: Then, readers can receive and process the information without needing a line of sight, and then create a report about the system [29]. Readers, also known as transceivers (transmitters/receivers), are devices designed for use with a radio frequency interface (RFI) module and control unit [32]. The main function of readers is to activate tags, interact with tags, and also exchange information between tags and software applications.

Readers can track the movement of the tags in real-time, and thus, the movement of the objects to which they are attached [29]. In the interaction technologies' operation, RFID tags and miscellaneous digital/analog sensors have been used as IoT technologies' foundation [33]. When using RFID, the chip is used to convey the data; it is attached to an object and uses a wireless link to transfer data. RFID can be implemented in an industrial area for production and management as well. Figure 5 indicates how technologies' aspects changed from 1980 to 2009. Over this period, IoT technologies affected information communication technology (ICT) and innovation [34].

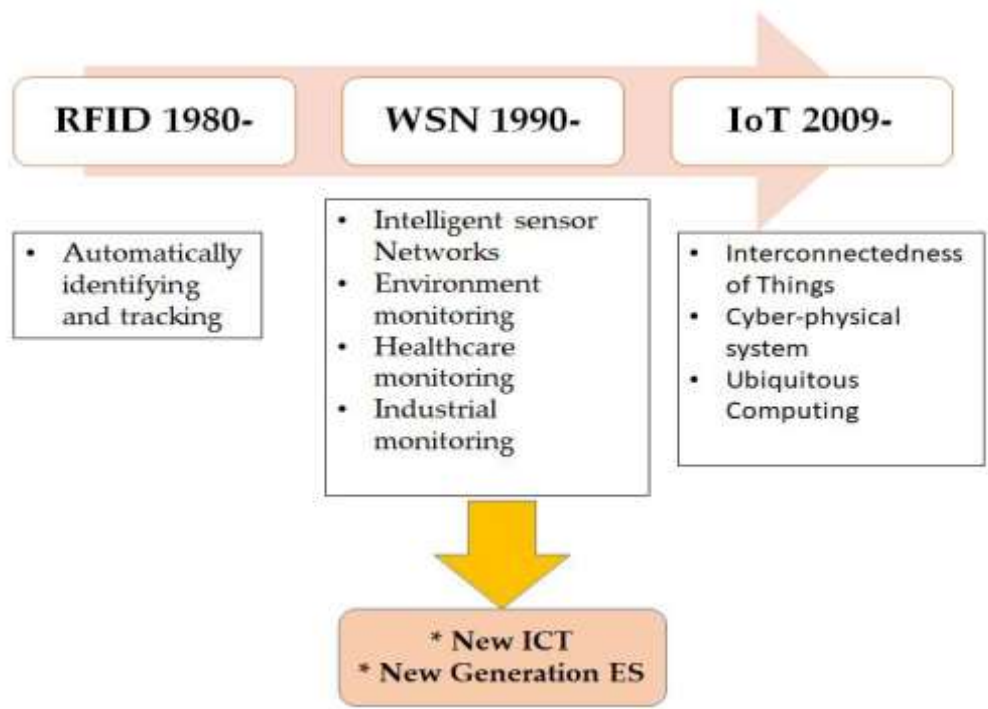

**Figure 5.** IoT-related technology and its importance for new ICT and enterprise systems.

2. Cloud Computing (CC)

The technology of cloud computing (CC) has offered many services, for example, Operating Systems (OSs) and application programming interfaces (APIs), for web applications. The success or failure of IoT projects is linked to big data (BD) because having a large number of resources in computing areas allows the CC paradigm to provide a new level of convenience in handling the BD generated by manufacturing IoT [29].

3. Wireless Sensor Network

The wireless sensor network (WSN) platform has been applied to the industry and built into IoT technologies. Using WSNs, wireless sensor and actuator networks (WSANs), virtual network sensors (VSNs), robots, machines, and gadgets, huge data can continuously be generated [35]. Recently, the technology for wireless communication and WSN systems has been dramatically researched, and many uses of WSNs have been identified, such as in industry, farming, healthcare, and many other fields. Industrial control, health monitoring, traffic control, and environmental monitoring are among the applications of WSNs [33,36,37]. In addition, some of the developed technologies such as ZigBee, LoRA, and LoRaWAN can be included in the WSN:

3.1. LoRa (long-range): For the IoT, the low power wide area network (LoRaWAN) protocol provides a long-range, low-power, low-bitrate communication method [38]. If you have a battery-powered device, this method works well. It has a long-range and relatively low power consumption [38,39].

The LoRaWAN is LPWAN (low-power wide-network technology) that recently gained a lot of attention because of its use in IoT network systems [39]. A large amount of data is transferred over long distances using this method. It is ideal for the WSN due to its high levels of sturdiness and coverage. The LoRa is a physical layer created by Semtech [38]. In the 250 bps to 5.5 kbps range, LoRaWAN data transfer rates are found. It employs bandwidths and frequencies between 125 and 500 kHz. Two types of devices exist in the technology: nodes and gateways. A node communicates with the gateway by sending and receiving data. Thousands of nodes can be connected to a gateway at once [38].

3.2. ZigBee: This is promoted by the ZigBee Alliance, which consists of hundreds of ZigBee member companies from semiconductors and software developers to original equipment manufacturers: Ember, Freescale, Chipcon, Invensys, Mitsubishi, CompXs, AMI AMI Semiconductors, ENQ Semiconductors. ZigBee and 802.15.4 are not the same [40].

ZigBee is a network protocol used as a standard data transfer protocol. The ZigBee Alliance, which is specified by the protocol and which enfolds Network Layer, Application Layer, and Security Layer, includes the physical layer (PHY and MAC—Media Access Control) and the network layers, which use IEEE 802.15.4 [41]. It can be noted that ZigBee-based wireless applications are increasingly demanded in most industries [40].

4.    Service-Oriented Architecture (SOA)

In research fields like cloud computing, WSNs, and vehicular network, the SOA has been utilized successfully [42,43]. A variety of ideas to build multi-layer SOAs for the IoT based on selected technologies, business needs, and technical specifications have been suggested [34]. The SOA can concentrate on the design and reuse of software and hardware components and enhance the workflows of the IoT architecture [27,34,44].

5.    Smartphones

In expanding and using this new IoT area, smartphones and other IoT devices will be crucial, as smartphones are widely accepted to be "at the heart of a growing universe of connected devices and sensors" [45]. The growth of smart wearables, like the Apple Watch and Android Wear, also assumes an important role in generating a smart body area network (BAN) for the user, who is mostly connected to them [45,46].

6.    Near Field Communication (NFC)

NFC guarantees that IoT technologies in a smart home are effectively implemented. The customer faces several tedious measures in linking devices without NFC in a smart home [47].

NFC is similar to RFID, but it provides improved protection and functionality [48]. An article published on the NFC Forum discusses four primary topics for which NFC is required in the IoT. These are presented as follows:

6.1    NFC can link any unpowered computer to the network. Useful information about an item can be obtained by tapping an NFC tag.

6.2    The user will pick their link using NFC, which offers the easiest and most convenient way for the user to proceed.

6.3    By adopting NFC, you can simplify the handshake process to connect devices with a simple tap.

6.4    The possibility of eavesdropping with NFC is greatly reduced.

Researchers from information, sociology, policy science, and computer sciences are included in infrastructure studies and a cross-disciplinary research field. The aim is to establish case studies and new methods for researching the role of infrastructure in society [30].

## 4. Advantages and Problems of the IoT

It is clear that every new technology has both positive and negative sides; the IoT, like any other new technology, has both advantages and disadvantages. The future with the IoT is radiant and bright but needs a lot of time and work and the route is hard. It is clear that every step at the starting point is difficult, but step by step, it will become easier and gradually most of the issues will be fixed. This is true for many new technologies, including the IoT, which, at least in the beginning, will encounter some difficulties that must be overcome. When it comes to the IoT, there are numerous technical issues to be resolved, as well as a lengthy process to gain a factual overview of it. The following two sections focus on the main advantages and disadvantages of the IoT:

### 4.1. IoT Advantages

According to [49], the following are some of the advantages of the IoT:

1. Technical enhancement:

IoT node usage and consumer observation of IoT facts can be enhanced by the use of similar techniques and data, which facilitates the most significant technological advancements. The real-world data performance and field performance can now be measured thanks to the IoT.

2. Enhanced consumer engagement:

There are ambiguity and fundamental errors in the precision of recent statistics and blockchain engagements in the IIoT remain inactive. The IoT transforms this, allowing for a rich and fruitful engagement that includes the spectator.

3. Advanced information compilation:

Today's information gathering is constrained by plans for practical use. The IoT slams it into those cracks and then places it right where people want to look at our globe.

4. Decreased waste:

The development fields generated by the IoT are more distinct. Instead of relying on statistics, the IoT provides us with real-world data that can help us better manage our resources.

Also, [50] outlined some other benefits of the IoT:

1. Improving technology:

Similar advancements and insights that enhance the client's experience also promote tool utilization and contribute to tremendous innovation over time. The IoT opens to a world of useful and field data [50,51].

2. Improved data collection:

Data collection for passive usage is innovative, but it has restrictions and designs. The IoT deconstructs it into areas and creates a space where people may go to evaluate our reality. It gives you a clear view of everything [50,52].

*4.2. IoT Problems*

In this section, some IoT problems are outlined:

1. Security and privacy:

In the IoT, technology systems are interconnected and communicate over networks. Given the fact that IoT is a network-based system, it is vulnerable to all sorts of cyber-attacks. The system provides detailed substantial personal data even when the user is not actively participating. The risk of losing privacy increases as more and more IoT data is transmitted. How secure will the data be in storage and transmission [53]?

2. Complexity:

The overall process of developing, designing, enabling, and maintaining IoT technologies is complicated for such an intelligent system.

3. Flexibility:

Consumers are concerned about the IoT scheme's ability to be easily integrated. The fear is that they will find themselves with too many conflicting or protected source codes [49].

4. Marketing and content delivery:

The IoT enhances this by observing additional habits and surprisingly deconstructing them. More data and information are generated as a result of this, leading to more accurate measurements and examples [50].

## 5. IoT Technologies' Architectural Design

To obtain accurate data from IoT applications, the architectural design must be programmed by the requirements of the IoT technologies. Therefore, choosing suitable hard-ware and the software type helps work with the IoT application type. The IoT technologies' architectural design has been founded, for instance, on web applications, smaBrt domains, and networking, among others.

Many types of research have been conducted on IoT technologies. Many reviews can be found that mention the IoT architecture, and some discuss different layers, for example, in [34,54], the authors indicate that the IoT architecture contains five layers: the sensing, accessing, networking, middleware, and application layers. But [55] outlines the five layers of the IoT architecture as: perception, network, middleware, application, and business (see Figure 6). Whereas, in [18,54], the authors state that the construction of the IoT includes three layers: the application, network, and sensing layers.

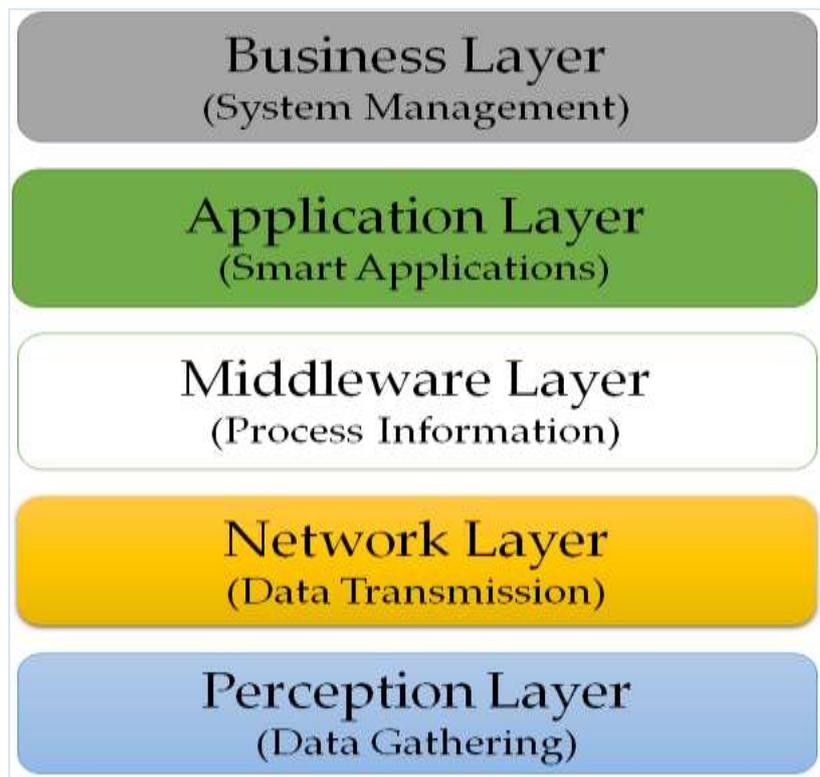

**Figure 6.** Five layers of IoT architecture.

According to [56], several researchers added an adaption layer to the IoT architecture, making it a six-layer system. This layer, which is between the perception and network levels, is an adapter that allows heterogeneous IoT devices to communicate with one another.

The IoT architectural designs are:

1. Sensing layer:

Perception or sensing layer [57] incorporates objects and data (RFID, sensors). These technological advancements vastly boost the IoT's ability to recognize and identify objects or environments. Intelligent service deployment techniques and universally unique identification (UUID) are assigned to each service or device that may be required in particular industry sectors [34].

This level includes objects that, owing to their ability to interact with the external world and their computing capabilities, become "intelligent" or "smart". Here, smart refers to the technological aspects (the smart technologies used), and intelligent refers to the functional aspects (self-identification, self-diagnosis, and self-testing), for example, sensors [58–60].

2. Networking layer:

This layer is responsible for specifying networking support and for transferring information operation via the wireless and wired networks.

A network layer can be attached to facilitate the transport of data from the perception layer to the processing layer [56]. The networking layer can collect data from current IT infrastructures (e.g., business systems, transportation systems, power grids, healthcare systems, ICT systems, etc.) [34].

It encompasses all of the technologies and protocols that enable this connection, and should not be mistaken for the International Organization for Standardization/Open Systems Interconnection (ISO/OSI) model's network layer, which just routes data within the network along the best path [58,59].

Wireless protocols are very significant at this layer. Wireless sensors, as opposed to those that require cables, can be deployed in difficult-to-reach areas and require less material and human resources to install. Furthermore, in a wireless sensor network, nodes can simply be added or withdrawn, and their placement can be modified without having to rethink the entire network's structure. The size of the network, the power consumption of each node, and the transmission speed required in a given application all influence which protocol is used [58].

3. Service layer:

The service layer (or application layer) [34] manages and creates services so the user can achieve with it what they require.

To ensure that the service can be delivered efficiently, the following enabling technologies should be incorporated in the service layer: Resource management and sharing technology, interface technology, service management technology, middleware technology, and interface technology [58]. Customers can request services ranging from agriculture to smart healthcare through the application layer [56,60].

Analytical tools like databases and analytical software are employed in the last stage of data processing and storage. It is through this processing that the data are made available for use by actual IoT applications (smart wearables, smart cars, etc.) [58]. This level also includes the control of the format of the data that are to be processed. Some of them can be classified as being of the following kinds [58,61]:

- Tiny, illegible to humans, binary-based data.
- Text-based, legible to humans, larger-format data.

Many firms are working on middleware service specifications, which are a major part of the service layer. Many common application requirements can be identified by a well-designed service layer [34].

4. Interface layer:

The work of this layer is to grant method services for the users, interface, and other applications.

Interface profiles (IFPs) can be considered as a subset of service standards that permit interaction with applications distributed on the network, and this layer is required to simplify the management and connectivity of things [34]. The implementation of Universal Plug and Play (UPnP), which provides a standard for promoting interoperability with services provided by diverse devices, is a key component of a decent interface profile [34,62,63].

## 6. Trends, Characteristics, and Application of the IoT

This section presents some of the trends and characteristics of the IoT, and then discusses some specific applications within the IoT area.

### 6.1. Trends and Characteristics of the IoT

1. Architecture:

As nearly a billion entities are joining together by the IoT, this will create a demand for a lot of traffic, and in turn, require larger storage capacities. Creating a new architecture is a very difficult project and requires considering all desired aspects, such as scalability, interoperability, interface, reliability, and QoS (Quality of Service). Regarding the design of the architecture of the IoT, the SOA is a key objective to develop in conjunction with the Internet and interfacing with a large choice of technologies' frames and linked networks.

2. Intelligence:

In the future, the IoT will become an open network and also non-deterministic in the structure of intelligence. Real-time analysis of data at the edge is critical to new IoT systems that leverage embedded intelligence (EI), connection, and processing capabilities for edge devices [64]. Hence, people can test and gather digital traces through the IoT, while also engaging with extensive smart options to learn more about life, social relationships, and communication in the environment.

3. Time considerations:

Regarding time in the IoT, which will no longer be used in a regular proportion but will be reliant on every unit, such as the system of information and the operation, the IoT will, as a result, be reliant on an enormous IT system.

4. Size considerations:

The size of the IoT would be approximately 50 to 100 trillion "things" if it had the capability to monitor and track all objects' motions. The average urban dweller is surrounded by between 1000 and 5000 potentially trackable objects [65].

*6.2. Applications of the IoT*

The goal or target of the IoT is to create a combination of the computer-generated and physical worlds in many areas. The IoT's applications can be used within many areas, for instance, home, city, agriculture, healthcare, automotive, and other domains as well. The mobility of nodes, strong security measures, a wide network range, and dense capacity are all characteristics of transportation, agriculture, and remote healthcare applications [66].

In some of the literature for this paper, IoT applications are categorized into certain domains. One of the features and goals of IoT technologies is improving the human quality of life. Improved quality of life for the end-user community and support for infrastructure and general-purpose operations are among the primary goals of IoT applications [67,68]. Some general IoT applications are listed in Table 1, such as manufacturing, smart environment, smart city, smart water, and smart houses. With its wide range of applications and services in various markets and industries, the IoT is considered to be a fast-growing, pioneering technology [67].

**Table 1.** In general, some of IoT's applications.

| Application Field | Area | Current Application |
|---|---|---|
| Manufacturing | Retail | Retail Center |
| Smart Environment and City | | City Sense Smart Parking |
| Smart Water | | GBROOS |
| Smart Houses | Security, Health | Smart Fridge, Smart Thermostat |

Previously, many different types of research concentrated on specific applications for the IoT. According to the review conducted for this paper, different types of IoT applications can be classified into several domains, with the following examples of some of those fields:

1. Systems of intelligent transportation and logistics

Sensors, actuators, and processing power are increasingly being incorporated into modern automobiles, trains, buses, and bicycles, as well as roads and rails [18]. For improving smart means of transport, such as vehicles, we can use wireless networks and a global positioning system (GPS) to monitor and control vehicles, as well as to make use of traffic data allowing us to control and conduct traffic. Within this intelligent system, via wireless networks, a huge number of vehicles can communicate with each other. An additional typical IoT-CPS-based application is smart transportation, also referred to as intelligent transportation systems, which integrates intelligent transportation management and control systems with communication networks and computing techniques to make transportation systems reliant, efficient, and protected [40,69]. Smart vehicles can efficiently perceive and share traffic data and schedule drivers' journeys with great efficiency, reliability, and safety. They can also share that data with other vehicles. Recent years have seen the development and testing of smart vehicles (such as Google's self-driving car, etc.) [40].

On the other hand, products can be founded from related information, timely and correctly, and therefore, the whole source of variable and complicated industries can respond shortly. Based on RFID and NFC, real-time information processing technology allows for the monitoring of almost every link in the supply chain, including commodity design, raw material procurement, production, transportation and storage of semi-products, and distribution and sale of finished goods [18]. Also, in general, today's vehicles have improved significantly when using communication and capabilities, thus, to improve these capabilities and share underutilized resources among vehicles in a parking lot or on the road, IoT technologies can be implemented [34]. For instance, by using IoT technologies, it would be easy to determine the vehicle's path, the location of the vehicle, and monitor it as well.

2. Smart monitoring and conducting (home, office)

Another aspect of using applications for IoT technologies is monitoring and conducting smart homes and the smart office. Use-cases involving the smart home and industry necessitate automation features such as monitoring and control, which require bitrates ranging from low to high for smart house applications and low to moderate for industrial ones [66]. Various houses and offices have been set up with sensors and monitoring to further control the structure of buildings, such as lights, heating, ventilation, and air conditioning (HVAC) systems, for example, as well as conducting surveillance to meet broader security requirements [70].

3. Healthcare

The medical sector is another area in which IoT technologies can be applied; one of the standard outlets for applications that incorporate the IoT is in the medical field, as the IoT has gradually gained popularity in the healthcare industry. Many applications can be used for healthcare, but the main applications with IoT technologies for smart medicine include the visualization of material management and digitization of medical information [71].

Telemedicine is an increasingly popular method of healthcare. It enables providers to provide care that is superior to many in-person treatment modalities [72]. An important issue in the aging population is providing adequate healthcare and assistance to disabled and elderly people. To provide and develop healthcare, an objective such as ICT is appropriate. Telemedicine has revolutionized medical care access. Also, telemedicine allows medical professionals, especially doctors, to care for patients from anywhere at any time.

There is a lot of gain from using IoT technologies that have been specified to healthcare sectors via their application, such as grouping objects and people (staff and patients), identifying and authenticating people, collecting data, and sensing on an automatic basis [18,73].

One of the greatest tasks is tracking, to identify the objects in the node. This involves both tracking real-time positions, such as patient-flow monitoring in healthcare, and tracking of movement using choke points, for instance, pin-pointing to particular zones [18]. Furthermore, assisted living scenarios are one of the most critical IoT healthcare applications. Sensors can be located in the health monitoring equipment, and then patients can use them easily [74]. One of the examples of this is the doctor, who from their office can note the rate of a patient's heart, which can be collected by sensors from time to time [34]; thus, doctors can monitor and contact their patients at home without the need for visiting the hospital. IoT applications have many benefits in healthcare for each patient, the hospital staff, and also for physicians [75]. Figure 7 shows how IoT applications and devices have been used in the healthcare area [76]. Lately, IoT technologies have even been used in remote monitoring for patients at home who have diabetes or hypertension. Several renowned companies have introduced facilities and products to IoT healthcare. For example, an Apple Watch has been produced with health features [77,78].

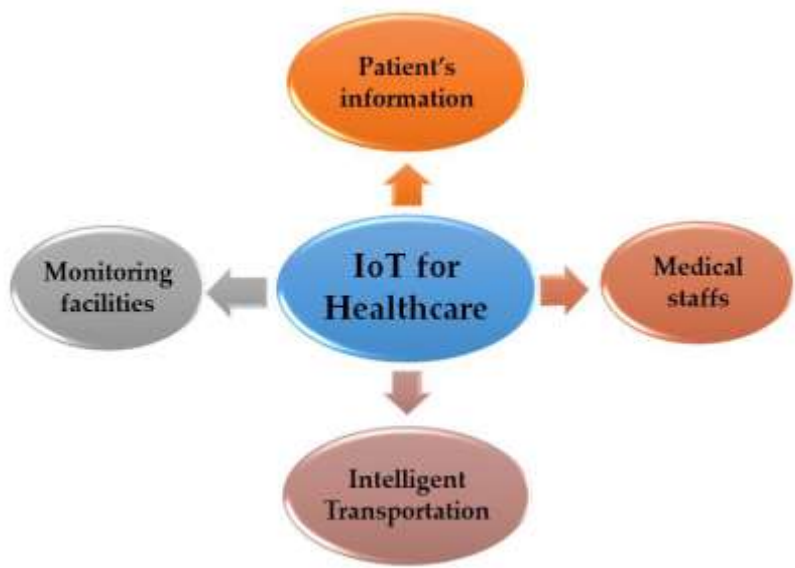

**Figure 7.** Internet of Things for healthcare.

4.  Personal and social

Devices can be connected in IoT technologies, and this is true even among humans; people can contact each other and exchange information, and they can build social relationships. Social networking is one of the common examples of IoT applications, for instance, using Facebook and Twitter. Additionally, IoT-enabled mobile phones can connect directly to other mobile phones and transfer contact information when predefined dating or friendship profiles are compatible with one another [74,79].

Another aspect of IoT applications is enabling people to use this technology to find things that are lost and search for their location as well. The search engine is one of the easiest web-based RFID applications that can be used to search for things in the last recorded location [18].

5.  Security and anti-theft systems

These features are now being advanced by incorporating the IoT to make the vehicle more cost-effective, secure, and dependable [4,80]. In addition to being easier to use, this technology is more reliable because it has features that can be controlled from a smartphone app to steer an embedded camera that can record the perpetrator's face as digital evidence [4].

6.  Monitoring system for pollution

Air pollution is important because it comes from cars, factories, and volatile chemicals. In this case, the transportation sector is a big source of air pollution. It contributes between 12% and 70% of air pollution [4,81]. The IoT can be used to keep an eye on the air pollution level in different places across the country [4,82]. Also, IoT technology can be used to measure the pollutants in the exhaust fumes of vehicles and send alerts to the owners and administrators of the vehicles, as well as suggest what else should be done [4,83].

7.  Environment monitoring systems

Another feature of applications of IoT technologies is their usefulness for environmental monitoring systems, such as for checking on the climate. Wireless nodes can be used to assess soil and also weather conditions. Also, within IoT applications, there are air pollution, waterway, and industry monitoring [84].

## 7. The IoT and Pandemic Control

In recent times, researchers from various fields around the world have focused on coronavirus (COVID-19) to find the best way to treat patients and develop a vaccine for it. A prime example is Apple's use of IoT technology in medical healthcare, particularly during the pandemic period. With COVID-19 patients, the IoT is expected to play a big role in the healthcare industry by collecting, processing, and analyzing people's symptoms through the IoT [10,85].

As a result of the pandemic, the number of IoT devices in healthcare applications has grown, resulting in an increase in data, which put a strain on the deployed healthcare IoT network in terms of communication metrics such as latency and throughput. This could lead to an incorrect diagnosis that could negatively impact the COVID-19 disclosure and service response rate [10,86].

Using elliptic curve cryptography (ECC) and a hash function, Leaby et al. came up with a way to make HC-IoT applications more secure in the case of a COVID-19 virus outbreak and emergency transportation [10].

Also, as mentioned previously, IoT technologies may be used to treat patients suffering from coronavirus infections. Currently, IoT technologies are used to determine when COVID-19 measures such as quarantine and mask-wearing are required [87]. The following are some of the ways in which IoT technologies can be used to face COVID-19:

1.  Using the IoT to examine a pandemic

One feature of IoT technologies is their capacity to examine a pandemic, through which patients can benefit from IoT mobile data used both upstream and downstream.

GIS (geographic information systems) can be used in this regard. Upstream uses can help identify those with COVID-19, while downstream uses can help isolate people who have come into contact with infected people [87].

2.   Remote health monitoring

Patients quarantined in a hospital or at home can be monitored consciously using this facility. Their medical examination is facilitated by hardware that prevents direct staff contact [88]. As mentioned previously in this paper, IoT technologies can be used to monitor patients, such as those with diabetes or hypertension, and this idea can also help patients with COVID-19. IoT technologies can also be used to reduce the healthcare workload and improve staff competency.

3.   Temperature sensor

Doctors must periodically check the patient's body temperature using various sensors such as thermistors and resistance temperature detectors (RTDs) [89]. An adult's normal body temperature is 98.6 °F (37 °C), but everyone's body temperature is slightly different, constantly fluctuating [90].

With COVID-19, a thermometer device (thermometer gun) is used to check the body temperature to see if it is higher or lower than normal (37 °C) [91].

4.   Using the IoT to ensure adherence to quarantine

The IoT can help collect data about all patients in quarantine. COVID-19 patients should be quarantined for 14 days, doctors recommend. For doctors and nurses, IoT data will help them check-in and monitor patients who have been quarantined [87].

The COVID-19 pandemic has had a huge impact on the IoT ecosystem [92,93]. Contact tracing and temperature screening are examples of uses that are directly tied to halting the spread of the virus, while others seek to assist the pandemic's new normal (e.g., working from home, homeschooling, home fitness, etc.) [92,94,95].

Technology influenced by the IoT has come to be used more widely and at a lower cost during the pandemic, such as the ADAMM Asthma Monitor, which detects an asthma attack before it happens [92,96], and other examples of how the IoT is being used in the healthcare sector include smart continuous glucose monitoring (CGM) [92].

It is possible that IoT applications, such as toll and reservation ticketing, navigation tracking, GIS mapping, guidance, and control systems for the supply chain as well as intelligent vehicles, can help alleviate some of the problems that may arise during a pandemic by limiting crowding in populated areas and speeding up import-export trade. Automation of roadways, trains, seaways, and airways can assist in eliminating human-to-human contact to lower the danger of infection, while allowing the flow of labor to continue [92,97].

The growth of modern pandemic prevention and control will be fueled by new IoT technology [98]. A lot of progress is being made in artificial intelligence (AI) and fifth-generation wireless (5G) technologies, which are making IoT applications more intelligent, more connected, and more powerful. This is because AI + IoT (AIoT) is becoming more powerful over time [98].

Some COVID-19 prevention and control techniques are highlighted below:

1.   Wearable Devices

Wearable technology, created at the Massachusetts Institute of Technology lab (MIT) from the 1960s, is a combination of multimedia, sensors, and wireless communication that is worn on the body [98]. Wearable technology is a combination of electronics and anything that can be worn, such as a watch or bracelet, for example [99,100].

These wearables are created to serve a variety of functions in a variety of fields, including healthcare, fitness, and fashion, among others [99,101,102]. It is anticipated that researchers will be able to monitor novel coronavirus pneumonia in the future. Wearable sensors offer a novel method of detecting COVID-19, which relies on long-term observation of changes in heart rate, breathing, coughing, and body temperature to determine the coronavirus infection [98].

Wearable gadgets are commonly utilized to monitor the health of potentially infected patients and self-detect physiological changes during isolation for COVID-19 management. Wearable gadgets detect a patient's whereabouts using GPS data, allowing clinicians to readily monitor their status [98].

2. However, the global system for mobile communications (GSM) is lagging behind the new era of 5G technology, which combines the GSM module with unique mobile apps and transmits information to and views all details in the mobile apps [98]. This is the only way to effectively monitor users. Using powerful digital signal processing methods, Fyntanidou's team developed a wrist-worn wearable device that continuously extracts vital signs such as the heart rate, oxygen saturation, and body temperature [98,103].

A wearable body sensor, network application programming interface layer, and mobile frontend layer are all part of Bassam's 3D wearable prototype design [98,104]. An automated healthcare system can also be built using a portable wearable device [98].

3. Kinsa, a smart thermometer, is another excellent home health monitoring solution. Fever and illness can be mapped by the gadget, allowing users to respond quickly to their current condition. In addition, it keeps track of a family's health and medical history, manages prescriptions, and sets up reminders, all while being child-friendly [92].

It is possible to save people's lives with IoT-based emergency medical services (EMS) solutions. In [92,105], the author proposes a system that delivers real-time data on the number of available beds, blood levels of various types, blood type availability, and doctors' availability. During big incidents or when there are several casualties, real-time data from the ambulance can be retrieved [92].

## 8. Industry 4.0 and the Industrial Internet of Things

In the last few years, the IoT has rapidly expanded as an overarching concept that is used both to use extended Internet technology as a global network that interconnects various artifacts and to incorporate such a vision into technology (including, e.g., RFIDs, sensors, protocols for physical and network layers, etc.) [106,107]. The IoT and IIoT, besides the digitalization of industrial manufacturing, can be considered as having instigated the fourth industrial revolution (Industry 4.0) [67].

Several industries use the IoT. The redesign of the industrial IoT model will speed up the advent of the next industrial revolution, also known as "Industry 4.0" [16,108,109].

One of the foremost characteristics of Industry 4.0 is the IIoT. The IIoT is supposed to be an excellent aspect of Industry 4.0, which will include highly efficient technology arrangements in many of the current industrial fields like transportation and manufacturing [109,110]. In addition, the IoT is a component of the Industry 4.0 concept, which emphasizes the idea of consistent digitalization and connectivity between all productive units, while still retaining the advantages of the traditional manufacturing process [110,111]. This section briefly reviews some points within Industry 4.0 and the IIoT. Some functions of Industry 4.0, the IIoT, and IoT are presented in Figure 8 [35].

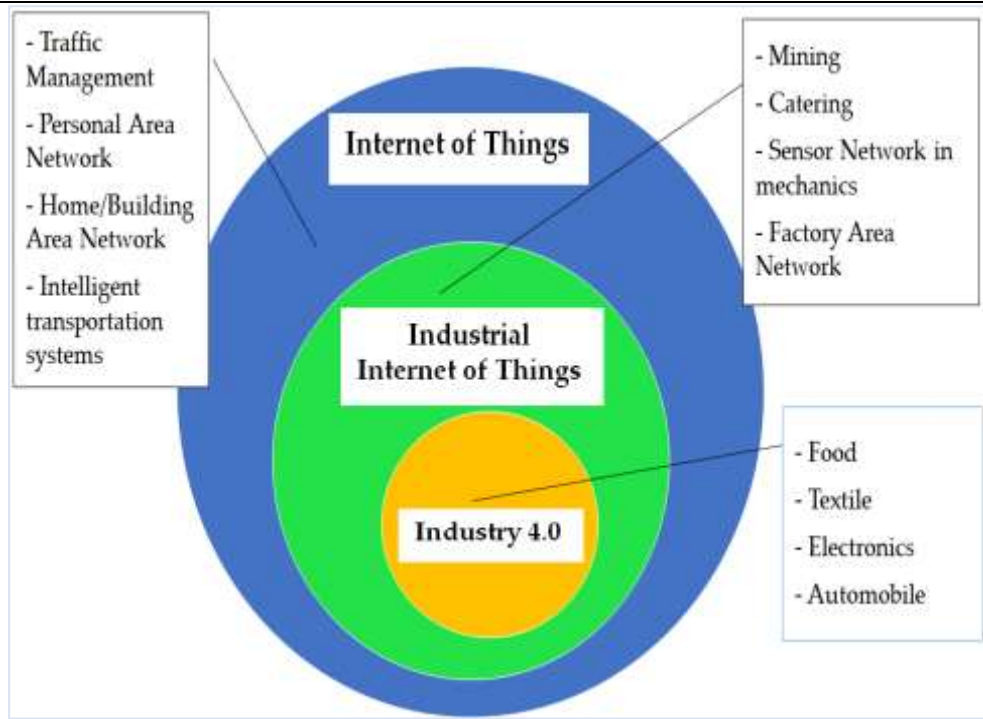

**Figure 8.** The IoT, IIoT, and Industry 4.0.

*8.1. Industry 4.0*

The term "Industry 4.0" refers to the fourth industrial revolution, which is characterized by the digitalization of the system's physical assets as well as the smart development of goods through the use of internet-based technology [112,113]. In the manufacturing fields, machines and sensors can be connected via the Internet to work better. Implementing Industry 4.0 would make devices and also sensors interconnected to the Internet; this makes the work easier. Table 2 indicates how the industry improved from Industry 1.0, from 1784 to the present, with Industry 4.0 incorporating IoT technologies [114].

**Table 2.** Industry 4.0.

| | Years | | | |
|---|---|---|---|---|
| | **1784** | **1870** | **1969** | **Today** |
| Industrial Revolution | Industry 1.0 Mechanization Steam power Weaving loom | Industry 2.0 Mass production Assembly line Electrical energy | Industry 3.0 Automation Computers and electronics | Industry 4.0 Cyber-physical systems Internet of Things networks |

Industry 4.0, which is a vital German ambition, intends to modify intelligent plant production technologies, transforming and upgrading these with cyber-physical systems (CPSs), the IoT, and cloud computing [115–117]. It is necessary to include the IoT as an enabling tool for connecting and cooperating humsn in real-time with cloud-enabled infrastructure and cyber-physical systems, in the ongoing development that is currently guided by Industry 4.0 [118].

With the integration of advanced manufacturing systems with intelligent manufacturing processes, Industry 4.0 enables a new era in technology that fundamentally alters the supply chains, value chains of production, and business models of industries [117]. Cloud services have helped to revolutionize unique IT cloud infrastructures that allow the representation of devices, which allow the incorporation of technological applications and AI bots [118–120].

Aside from sensors, smart factories have successfully adopted many RFID applications [121–123]. This technology is critical for improving a variety of industrial automation tasks, including automatic identification of products, instruments, and other production tools, as well as inbound/outbound tool management [121,124,125].

The following examples will demonstrate the value of diverse and widespread uses in the automobile industry:

1. Traffic management and toll collection system

The traffic management system plays an important part in ensuring the nation's economic stability [4,126]. With the increasing number of vehicles on the road, intelligent traffic signal timing depending on vehicle density can be managed using the IoT, locally deployed RSUs, and security cameras [4,127,128].

It informs the driver about the road condition (e.g., asphalt, wet, snow), traffic situation, and other unexpected information on the route, as well as emergency parking options [4,97].

2. Real-time vehicle navigation

An IoT-based real-time car navigation system has a huge impact on guaranteeing a safe and efficient transportation system [4,129].

When a vehicle encounters emergency issues such as an accident, an IoT-enabled model employing the SKM53 GPS module and the Haversine formula alerts the nearest rescue team for immediate assistance and speedy aid to victims [4,130]. The IoT applications can also be used to track school buses [4,131].

## 8.2. Industrial Internet of Things

The adoption of IoT technologies by the industry, the definition implemented by the German government in 2011 for the strategic planning of their manufactures, is also called IIoT or Industry 4.0 [106,132], in which today the new revolution of this industry is in use. To manufacture tangible products for the marketplace as well as to maintain the physical properties related to production, the Industrial IoT (or IIoT) should be understood as part of the larger IoT [57]. The IIoT, in smart manufacturing systems, will be able to produce a revolution in the process industry [57]. The IIoT can be considered one of the specific types of IoT; it is focused on its applications and use cases in modern industries and smart manufacturing, rather than on the Internet in general [67].

The IIoT, in a manufacturing firm, is undeniably empowering the industrial revolution (Industry 4.0) [133]. The IIoT differs from the customer IoT in that it focuses on interfacing smart gadgets in the fundamental framework, assembly, and procedure enterprises to create business points of interest [57,134]. The IIoT can be used for IoT technologies in the manufacturing sector.

An IIoT environment cannot be complete without WSNs and WSANs, and middleware that can monitor them [35]. More recently, the IIoT has emerged as an inferior model focused mainly on safety-critical applications in industrial operational applications [34,135]. IIoT technology and data analytics utilize (record, monitor, manage) real-time data to facilitate and illuminate money-saving choices [135]. In Figure 9, a proposed architecture for the manufacturing industry IIoT is demonstrated [54].

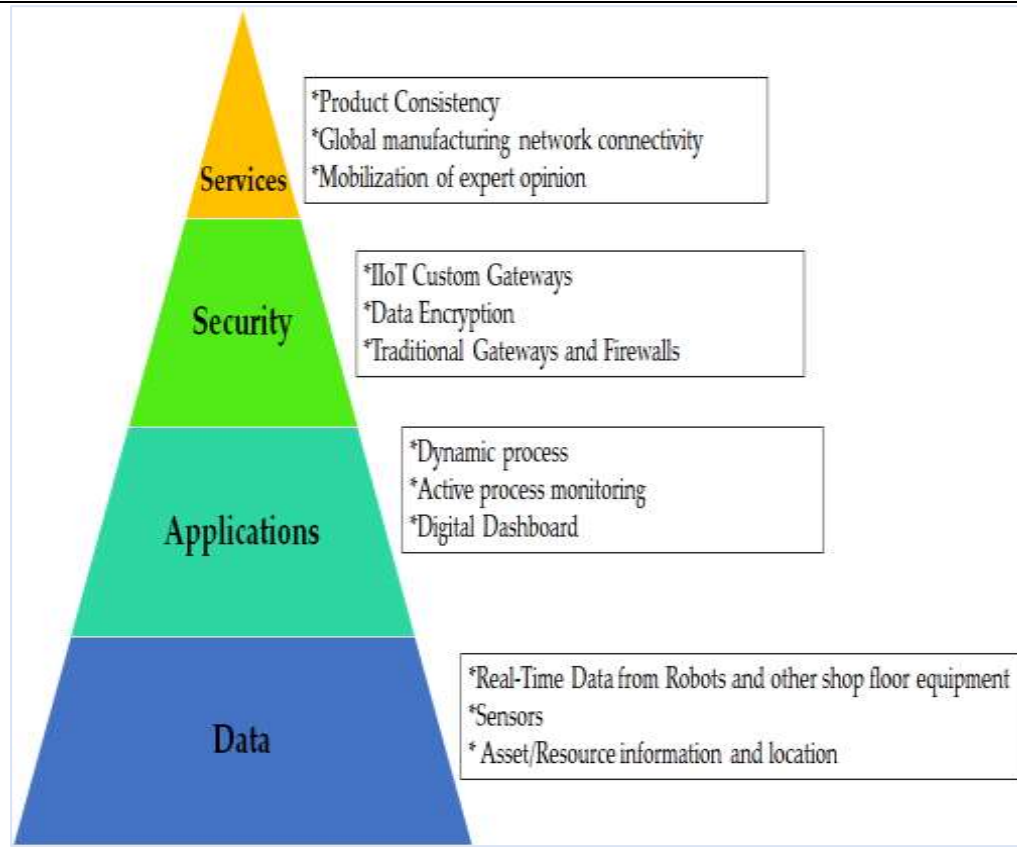

**Figure 9.** The proposed potential Industrial IoT (IIoT) architecture for the manufacturing industry.

According to [136], the IIoT connects many points of tools, data, and people throughout industrial environments, and it provides numerous major benefits once established.

1. Automated Production

A key advantage that IIoT technology provides to its users is the ability to automate certain processes. Automated systems can be trained to conduct manual tasks that would normally be done by human workers on an assembly line using the IIoT. Automation in the IIoT reduces the amount of manual work that must be done by qualified personnel.

2. Maintenance and Safety

Automated predictive maintenance and safety monitoring in manufacturing lines are common uses of the IoT. IIoT sensors provide firms with the capacity to examine numerous elements of performance and identify whether equipment needs to be updated or replaced or whether a worker is in contact with harmful working circumstances. For the protection of goods in transit, IIoT sensors may also monitor environmental elements like temperature and air quality.

3. Real-Time Efficiencies

It is common to utilize the IIoT as a preventative measure to avoid downtime due to equipment failures and other performance concerns because of its emphasis on gathering as many real-time data points and insights as possible. Having less downtime in the workplace leads to improved productivity and efficiency in the workplace.

4. Workforce-Equipment Connectivity

Human workers run equipment in a more traditional factory or industrial setting, while automated machines follow their programming. The IIoT is one of the few technological solutions that breaks down barriers between workers and their equipment. Users acquire more direct insights from their tools, and tools learn over time as a result of human engagement.

## 9. IoT Marketplace

This section reviews some of the trends and facts related to the IoT technologies in the marketplace, and how the population has been affected in this area. Firstly, global pollution must be taken into consideration as this subject is directly linked with technology areas, especially nowadays.

It is undeniable that the population has increased throughout the world and that it will continue to grow at an alarming rate. Humans, devices, and "things" are all connected in the technology world, and population growth has had an impact on the number of interconnectors available, including those in the IoT technology domain as well. By 2020, nearly every person on the planet had more than six devices connected to the Internet, according to a report from Cisco [137], which explains how the world population has grown in comparison to the number of connected devices per person over the period from 2003 to 2020, and demonstrates how the number of connected devices will increase dramatically. It was predicted that the number of devices that could be interconnected with one another would reach nearly 50 billion by 2021, which represents more than three times the number of devices that were connected in 2010 when 12.5 billion devices were connected [137]. The population has a significant and direct relationship with new technologies, including the IoT, and this has had an impact on the number of smart devices that can be interconnected with one another on the one hand, and on the ability of the IoT to influence the marketplace on the other.

IoT technologies have emerged as one of the most important aspects of the environment's domain; through the use of IoT technologies, millions of devices can communicate with one another and exchange information. As reported by Cisco [137], the total number of Internet-connected devices is expected to exceed 50 billion by 2020. According to a report by [138], one trillion networked sensors will be embedded in the world around us by 2022, with up to 45 trillion in 20 years.

While it is true that the number of interconnectors has increased as a result of new technologies, it should be noted that IoT technologies are also included in this topic. Using the infrastructure provided by the IoT domains, a massive number of devices and things can be connected to exchange information among themselves.

Aside from that, according to [139], the comprehensive IoT and the number of interconnected devices within IoT areas is expected to grow significantly from year to year, reaching 75.44 billion globally by 2025, representing a five-fold increase in ten years. Figure 10 shows IoT-connected devices were and are predicted to be installed worldwide from 2015 to 2025 (in billions) [139].

The IoT has advanced in a variety of areas over the years, and it is expected to continue to grow in the future. IoT technologies have become one of the most significant revolts in the digital domains on the commercial marketplaces for a variety of goods. Recently, the IoT has had an impact on a variety of areas in human life; it is clear that the IoT is a gateway to the marketplaces that is currently trending, and that IoT technologies have had an impact on the majority of marketplace sectors. Numerous factors have had an influence and are being tracked, including the growing environment, management, production, transportation, and markets, among others.

Because Microcontroller for IoT technologies has been thought by many companies from many countries, the trends of IoT technologies are growing in the field of marketplaces. Various devices can be used within the facility of the IoT marketplaces, and as mentioned by [140], some of the IoT casting devices in the market are the Apple Airplay, Google Chromecast, and Amazon Firesticks.

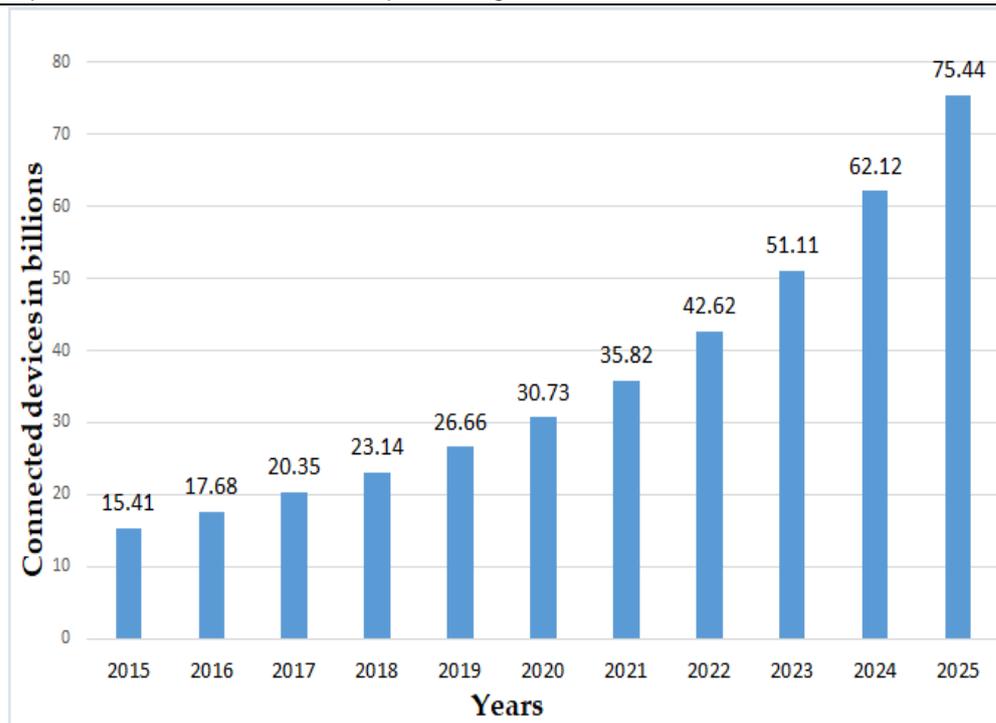

**Figure 10.** IoT connected devices installed worldwide from 2015 to 2025 (in billions).

In the IIoT sector, there is an increasing number of IT businesses, and some of the best providers have made the most progress: ABB, Cisco, Huawei, Intel, etc. [136]. This is where the vast majority of IIoT use-cases are now being implemented. To improve their operational visibility and real-time data understanding, businesses are turning to leading IIoT providers.

A joint initiative between Huawei and DHL will focus on cell-based IoT technology that can link high numbers of devices over great distances with little power usage. Data and insight into warehousing operations, freight transportation, and last-mile delivery will be provided by enhanced connectivity, which will result in a more integrated logistics in the value chain [136,141].

[136,142] intended to collect real-time quality data and use the same data to identify trends that could help prevent faults before they occur. The researchers' goal was to keep production moving as quickly as possible by alerting the person in charge promptly if any of the equipment failed or there was a power loss.

Table 3 shows the IoT endpoint market segmented by industry, between 2018 and 2020, with segments spanning utilities and governance, to building, automation, and physical security, to manufacturing and natural resources, to automotive and healthcare providers, to retail and wholesale trade, to information and transportation [143].

**Table 3.** IoT endpoint market by segment, 2018–2020, worldwide (installed base, billions of units). Source: Gartner (August 2019).

| Years \ Segment | Utilities | Government | Building Automation | Physical Security | Manufacturing & Natural Resources | Automotive | Healthcare Providers | Retail & Wholesale Trade | Information | Transportation | Total |
|---|---|---|---|---|---|---|---|---|---|---|---|
| 2018 | 0.98 | 0.4 | 0.23 | 0.83 | 0.33 | 0.27 | 0.21 | 0.29 | 0.37 | 0.06 | 3.96 |
| 2019 | 1.17 | 0.53 | 0.31 | 0.95 | 0.4 | 0.36 | 0.28 | 0.36 | 0.37 | 0.07 | 4.81 |
| 2020 | 1.37 | 0.7 | 0.44 | 1.09 | 0.49 | 0.47 | 0.36 | 0.44 | 0.37 | 0.08 | 5.81 |

Wireless radio frequency identification technology is one of the techniques that can be used to connect everyday objects to computer networks [16,31]. It is anticipated that RFID will continue to be the driving force behind the IoT, according to [144,145]. RFID has the potential to transform dump devices into objects that are comparatively intelligent [16]. According to [146], the retail segment of the RFID market will have grown from $2.7 billion in 2018 to $5.4 billion by 2021, representing a compound annual growth rate of 38.9 percent over the forecast period.

RFID has dominated the marketing strategy since its inception, owing to its low cost and small size, as well as its widespread use [145]. Economic fluctuation is another significant challenge impeding the more active development of the RFID market, particularly in market areas such as smart healthcare, smart cities, and smart transportation [146]. RFID can be used in manufacturing for a variety of purposes, including supply chain management, production planning, and more [29]

The IIoT is mostly used in manufacturing environments where automation plays an important role in success, but the technology is rapidly expanding in both capabilities and application cases [136].

For industrial operations, the IIoT market alone is developing at a fast rate despite being just one aspect of the overall IoT industry. According to [147], with a market size of roughly $216.13 billion in 2020, the worldwide IIoT market will reach about $1.1 trillion by 2028 [136].

The IoT is widely recognized as a cutting-edge technology that has the potential to transform the industrial sector [29]. By using advanced technology for knowledge and production, intelligent manufacturing can create manufacturing processes that are scalable, efficient, and easily reconfigurable to meet the demands of a globally competitive market [117,148]. Technology such as the IoT can enable manufacturing, as demonstrated by the German Industry 4.0 initiative, which has been influenced by the global economic environment in general.

As with a developing marketplace, IoT technologies have expanded to become a global market for the acquisition of illustrious products such as Google glasses and GPS shoes. A wireless sensor network, on the other hand, makes it possible for purchasers to access an electronic market from any location at any time. This is made possible by the facility of the electronic market. The field of wireless communications has seen significant advancements recently, particularly in the area of remote sensor networks, which have developed quickly and have been successfully implemented in the buyer electronics market [149], once again, this has become a global market, particularly in light of the ability of IoT technologies to be deployed in the marketplace.

According to IDC, the IoT solutions are at the heart of the third platform's vision. In recent years, the IoT solution has been introduced into the marketplace through a variety of techniques; as a result, these solutions can be implemented both economically and creatively, thus introducing the concept of the third platform. Figure 11 describes innovative industry solutions and the IoT driving value throughout the third platform [150].

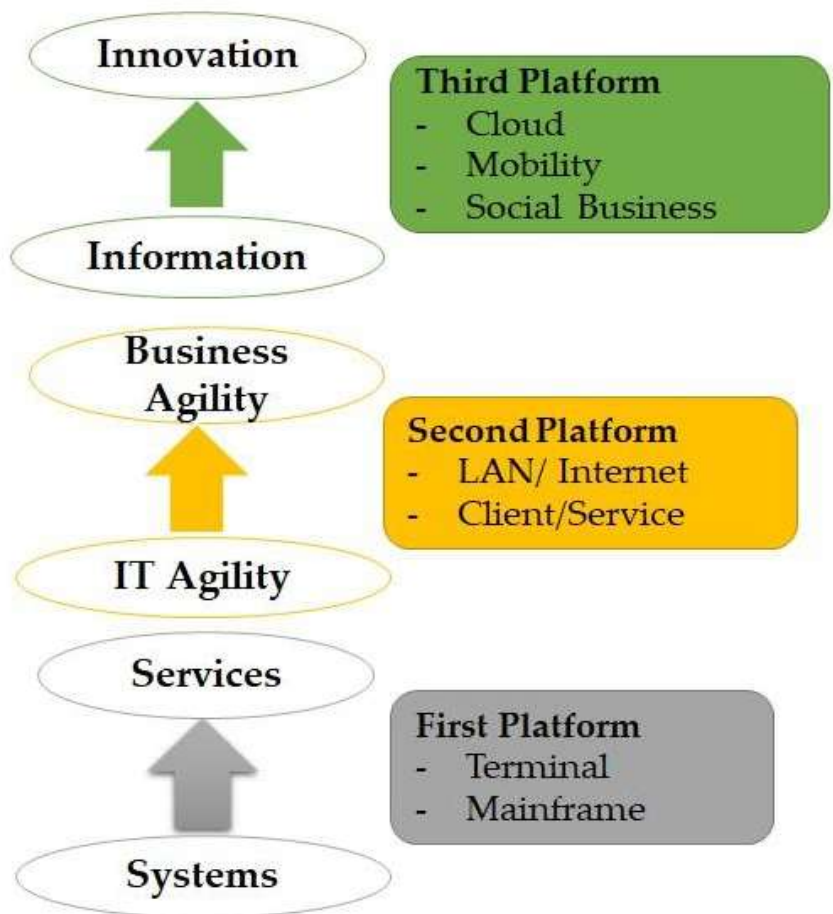

**Figure 11.** IoT: Driving value throughout the third platform.

*IoT Development Market: Exports of Connected Devices*

As indicated by [25], the growth of the IoT marketplace has been driven by communication among devices. Consistent with [151], the IoT development market has seen growth in different periods from 2015, and that it is expected to continue up to 2030, while two points have been discussed concerning commercial and industrial electronics and consumers as well.

According to [151], in general, the description of the IoT marketplace has been highlighted as below:

1. Commercial & Industrial Electronics

   • Good growth is projected for sales: 2013–2030 compound annual growth rate (CAGR) = 20%.
   • The next decade (2021–2030) adds the highest amount of devices deployed.

2. Consumers

   • In the long term, revenue would see moderated growth: 2013–2030 = 13.8% (CAGR).
   • The user base increased from 2015 to 2020 for connectable devices.
   • New shipping for total compatibility will rise by 12.5% from 2013–2030.

The primary scheme of an IoT business opportunity is to allow the owners of sensors and actuators to charge IoT application designers for the use of sensor-generated data and access to actuator capabilities. This type of commerce goes beyond the immediate needs of device owners and application engineers, allowing for the development of a new class of data-driven service contributors, where data processing and analysis can be provided to vendors and purchasers of IoT data by dealers, rather than the immediate needs of device owners and application engineers.

According to [151], an intelligent IoT integrator (I3) must incorporate the following components to be a practical and successful commercial center:

1. The dealers post their accessible sensor-flowing information and actuators admit items alongside some hint of their worth and states of utilization.
2. Application designers and data/actuators admit purchasers to peruse or be suggested pertinent sensor data streams/device proprietors just as data intermediaries to interface with. This can be an index or potentially an endorsement instrument.
3. Motivating mechanisms for vendors to afford valuable data and actuators, for example, adaptation per unit amount of data for the actuator to acquire.
4. Directing ongoing data from sensor data sources to all purchasers that are approved dependent on the established understandings (counting installments, procedure arrangements).
5. Steering continuous activation and control signals from approved clients (purchasers) to actuators. Assurance and prestige evaluating and determining components for clients to rate each other. Metering and charging instruments to uphold advertising agreements being steered through the commerce.
6. APIs and SDKs for methodical admittance to different functionalities of the platform data naming and organizing capacities, just as security instruments to channel data.

The IoT marketplace consequently presents: proficient sharing/exchanging of data, advantageous arrangement and exchanges of data and access cost and utilization strategy models, and dataflow control to actualize the business perception [151].

The research of [152] has explained some examples of IoT marketplaces, including:

1. For the Internet of Vehicles (IoVs) to function, monetary transactions must be made between vehicle-to-vehicle networks, vehicle-to-roadside networks, and vehicle-to-infrastructure and pedestrian networks.
2. Smart Agriculture—Real-time crop information is available to farmers and other agricultural supply chain participants using this technology.
3. Smart Health—Data monetization is possible thanks to the wearable IoT (WIoT) devices used by patients, medical imaging, test results, social media, and external patient data that are collected by healthcare systems and patients alike.

## 10. Conclusions and Future Work

The IoT has emerged as a significant point in most people's lives because it provides numerous benefits to them as a result of the numerous "things" that can be used to communicate and exchange information. However, just like any other technology, the IoT has its advantages and disadvantages, some of which were discussed in this paper.

Based on the findings of this paper, we can reap significant benefits from the application of IoT technologies across a wide range of sectors and environments, including intelligent transportation and logistics, healthcare, smart cities, and social services. Another aspect of this paper is concerned with the use of the IoT for telemedicine, including how some IoT technologies can be used to combat COVID-19 for patients who are infected with this new and deadly virus, and how IoT technologies can be one of the great gates for assistance in medical environments. Using IoT during the pandemic, and determining how to control it and use it in the future, is another important feature that has been discussed, with readers now more familiar with the technologies used in this era. Furthermore, certain methods that can be used to prevent and control COVID-19 have been highlighted. Thus, novel research can focus on this idea more and work on it in future studies. Another important part of this paper was the application of the IoT in industries, where Industry 4.0 and the IIoT are important points to consider. Crucial benefits of the IIoT have been clarified as automated production, maintenance and safety, real-time efficiencies, and workforce-equipment connectivity, among others. The IoT has been applied to the marketplace and has helped to brighten it; major points have been discussed in this paper, such as IoT casting devices currently available on the market and the IoT endpoint market by segment. Within this study, the IoT development market has been discussed to highlight the growth of the IoT marketplace for communication among devices. In the field of the IoT, numerous tasks have been completed; however, there are still additional tasks that must be completed in the future. In this study, various subjects related to IoT technology have been discussed, and so this article is helpful for researchers who are starting to work in this area. This paper used many references, which is again helpful for researchers who would like to become more familiar with the IoT field. Crucially, as healthcare is one of the most important areas for every human, the IoT has the potential to be used more in the medical sector in the future. IoT technologies have a broad range of applications and it is crucial that we integrate them with other technologies so that the applications and facilities for IoT healthcare can be improved further and to make them more accessible to all users.

**Author Contributions:** Conceptualization, K.O.M.S. and T.A.R.; methodology, K.O.M.S. and T.A.R.; formal analysis, K.O.M.S.; investigation, K.O.M.S. and T.A.R.; writing—original draft preparation, K.O.M.S.; writing—review and editing, K.O.M.S., T.A.R., D.R. and N.B.; supervision, T.A.R.; funding, T.A.R., D.R. and N.B. All authors have read and agreed to the published version of the manuscript.

**Funding:** This research work received no funding.

**Data Availability Statement:** Data and materials are available upon request.

**Acknowledgments:** The authors would like to thank the chief editors and reviewers for processing this manuscript. Special thanks go to the University of Sulaimani and the University of Kurdistan Hewler for proving all the facilities needed for this research work.

**Conflicts of Interest:** The authors declare no conflict of interest.

## References

1. Tan, L.; Wang, N. Future Internet: The Internet of Things. In Proceedings of the 3rd International Conference on Advanced Computer Theory and Engineering, ICACTE 2010, Chengdu, China, 20–22 August 2010; pp. V5-376–V5-380. https://doi.org/10.1109/icacte.2010.5579543.
2. Ashton, K. Internet of Things. Page 1—RFID Journal. Available online: https://www.rfidjournal.com/articles/view?4986 (accessed on 2 March 2020).
3. Okano, M.T. IOT and industry 4.0: The Industrial new revolution. In Proceedings of the International Conference on Management and Information Systems, ICMIS 2017, Bangkok, Thailand, 25–26 September 2017. Available online: https://www.researchgate.net/publication/319881057_IOT_and_Industry_40_The_Industrial_New_Revolution (accessed on 17 September 2021).
4. Rahim, M.A.; Rahman, M.A.; Rahman, M.M.; Asyhari, A.T.; Bhuiyan, M.Z.; Ramasamy, D. Evolution of IOT-enabled connectivity and applications in Automotive Industry: A Review. *Veh. Commun.* **2021**, *27*, 100285. https://doi.org/10.1016/j.vehcom.2020.100285.
5. Sharma, S.; Kaushik, B. A survey on Internet of vehicles: Applications, security issues and solutions. *Veh. Commun.* **2019**, *20*, 100182.
6. Silva, C.; Silva, F.; Sarubbi, J.; Oliveira, T.; Meira, W.; Nogueira, J. Designing mobile content delivery networks for the Internet of vehicles. *Veh. Commun.* **2017**, *8*, 45–55.
7. Lei, T.; Wang, S.; Li, J.; Yang, F. A cooperative route choice approach via virtual vehicle in iov. *Veh. Commun.* **2017**, *9*, 281–282.
8. Bajaj, R.; Rao, M.; Agrawal, H. Internet of things (iot) in the smart automotive sector: A review. *IOSR J. Comput. Eng.* **2018**, *9*, 36–44.
9. Fantian, Z.; Chunxiao, L.; Anran, Z.; Xuelong, H. Review of the key technologies and applications in internet of vehicle. In Proceedings of the 2017 13th IEEE International Conference on Electronic Measurement & Instruments (ICEMI), Yangzhou, China, 20–22 October 2017; pp. 228–232.
10. Adil, M.; Khan, M.K. Emerging IOT applications in Sustainable Smart Cities for COVID-19: Network security and data preservation challenges with future directions. *Sustain. Cities Soc.* **2021**, *75*, 103311. https://doi.org/10.1016/j.scs.2021.103311.
11. Mbunge, E. Integrating emerging technologies into COVID-19 contact tracing: Opportunities, challenges and Pitfalls. *Diabetes Metab. Syndr. Clin. Res. Rev.* **2020**, *14*, 1631–1636. https://doi.org/10.1016/j.dsx.2020.08.029.
12. Otoom, M.; Otoum, N.; Alzubaidi, M.A.; Etoom, Y.; Banihani, R. An IOT-based framework for early identification and monitoring of COVID-19 cases. *Biomed. Signal Processing Control* **2020**, *62*, 102149. https://doi.org/10.1016/j.bspc.2020.102149.

13. Vedaei, S.S.; Fotovvat, A.; Mohebbian, M.R.; Rahman, G.M.; Wahid, K.A.; Babyn, P.; Marateb, H.R.; Mansourian, M.; Sami, R. COVID-safe: An IOT-based system for Automated Health Monitoring and surveillance in post-pandemic life. *IEEE Access* **2020**, *8*, 188538–188551. https://doi.org/10.1109/access.2020.3030194.

14. Gillis, S.A. What Is Internet of Things (IoT)? Available online: https://internetofthingsagenda.techtarget.com/definition/Internet-of-Things-IoT (accessed on 28 December 2021).

15. Sundmaeker, H.; Guillemin, P.; Friess, P.; Woelfflé, S. In *Vision and Challenges for Realising the Internet of Things*; European Commission Information Society and Media: Brussels, Belgium, 2010. Available online: http://www.internet-of-things-research.eu/pdf/IoT_Clusterbook_March_2010.pdf (accessed on 7 November 2021).

16. Perera, C.; Liu, C.H.; Jayawardena, S.; Chen, A.M. A Survey on Internet of Things from Industrial Market Perspective. *IEEE Access* **2014**, *2*, 1660–1679. https://doi.org/10.1109/access.2015.2389854.

17. Lampropoulos, G.; Siakas, K.V.; Anastasiadis, T. Internet of Things (IoT) in Industry: Contemporary Application Domains, Innovative Technologies and Intelligent Manufacturing. *Int. J. Adv. Sci. Res. Eng.* **2018**, *4*, 109–118. https://doi.org/10.31695/ijasre.2018.32910.

18. Atzori, L.; Iera, A.; Morabito, G. The Internet of Things: A survey. *Comput. Netw.* **2010**, *54*, 2787–2805. https://doi.org/10.1016/j.comnet.2010.05.010.

19. ITU Internet Reports 2005: The Internet of Things. Available online: https://www.itu.int/osg/spu/publications/internetofthings/ (accessed on 10 March 2021).

20. Weber, R.H. Internet of Things–New security and privacy challenges. *Comput. Law Secur. Rev.* **2020**, *26*, 23–30.

21. Bairagi, J.; Joshi, S.; Barshikar, S. A Survey on Internet of Things. *Int. J. Comput. Sci. Eng.* **2018**, *6*, 492–496. https://doi.org/10.26438/ijcse/v6i12.492496.

22. Chen, S.; Xu, H.; Liu, D.; Hu, B.; Wang, H. A Vision of IoT: Applications, Challenges, and Opportunities with China Perspective. *IEEE Internet Things J.* **2014**, *1*, 349–359. https://doi.org/10.1109/jiot.2014.2337336.

23. Atzori, L.; Iera, A.; Morabito, G. Understanding the Internet of Things: Definition, potentials, and societal role of a fast evolving paradigm. *Ad Hoc Netw.* **2017**, *56*, 122–140. https://doi.org/10.1016/j.adhoc.2016.12.004.

24. Radoglou Grammatikis, P.I.; Sarigiannidis, P.G.; Moscholios, I.D. Securing the Internet of Things: Challenges, threats and solutions. *Internet Things* **2019**, *5*, 41–70. https://doi.org/10.1016/j.iot.2018.11.003.

25. Morelli, B. IoT Market Overview. 2018. Available online: https://www.gsaglobal.org/wp-content/uploads/2019/05/Bill-Morelli-GSA-Silicon-Summit-IoT-Market-Overview-IHS-Markit-4.9.18.pdf (accessed on 17 September 2021).

26. Li, Y.; Hou, M.; Liu, H.; Liu, Y. Towards a theoretical framework of strategic decision, supporting capability and information sharing under the context of Internet of Things. *Inf. Technol. Manag.* **2012**, *13*, 205–216. https://doi.org/10.1007/s10799-012-0121-1.

27. Haldikar, A.; Lalwani, P.; Pandey, S.; Chitari, A. IOT Based Industrial Management. *Int. J. Res. Appl. Sci. Eng. Technol.* **2017**, *5*, 2945–2947. https://doi.org/10.22214/ijraset.2017.11406.

28. Juels, A. RFID security and privacy: A research survey. *IEEE J. Sel. Areas Commun.* **2006**, *24*, 381–394. https://doi.org/10.1109/jsac.2005.861395.

29. Yang, C.; Shen, W.; Wang, X. Applications of Internet of Things in manufacturing. In Proceedings of the IEEE 20th International Conference on Computer Supported Cooperative Work in Design, CSCWD 2016, Nanchang, China, 4–6 May 2016. https://doi.org/10.1109/cscwd.2016.7566069.

30. Kranenburg, R.V.; Anzelmo, E.; Bassi, A.; Caprio, D.; Dodson, S.; Ratto, M. The Internet of Things. In Proceedings of the 1st Berlin Symposium on Internet and Society, Berlin, Germany, 25–27 October 2011. Available online: http://www.pyfn.com/PDF/iot_pdfs/the_iot_paper_2011.pdf (accessed on 10 April 2020).

31. Welbourne, E.; Battle, L.; Cole, G.; Gould, K.; Rector, K.; Raymer, S.; Balazinska, M.; Borriello, G. Building the Internet of Things Using RFID: The RFID Ecosystem Experience. *IEEE Internet Comput.* **2009**, *13*, 48–55. https://doi.org/10.1109/mic.2009.52.

32. Jia, X.; Feng, Q.; Fan, T.; Lei, Q. RFID technology and its applications in Internet of Things (IoT). In Proceedings of the 2nd International Conference on Consumer Electronics, Communications and Networks, CECNet 2012, Yichang, China, 21–23 April 2012; https://doi.org/10.1109/cecnet.2012.6201508.

33. Ahmad, J.; Chaudhary, J.; Ahmad, M.; Naz, A. Survey on Internet of Things (IoT) for Different Industry Environments. *Ann. Emerg. Technol. Comput.* **2019**, *3*, 28–43. https://doi.org/10.33166/aetic.2019.03.004.

34. Xu, L.D.; He, W.; Li, S. Internet of Things in Industries: A Survey. *IEEE Trans. Ind. Inform.* **2014**, *10*, 2233–2243. https://doi.org/10.1109/tii.2014.2300753.

35. Aazam, M.; Zeadally, S.; Harras, K.A. Deploying Fog Computing in Industrial Internet of Things and Industry 4.0. *IEEE Trans. Ind. Inform.* **2018**, *14*, 4674–4682. https://doi.org/10.1109/tii.2018.2855198.

36. Li, S.; Xu, L.D.; Wang, X. Compressed Sensing Signal and Data Acquisition in Wireless Sensor Networks and Internet of Things. *IEEE Trans. Ind. Inform.* **2013**, *9*, 2177–2186. https://doi.org/10.1109/tii.2012.2189222.

37. He, W.; Xu, L.D. Integration of Distributed Enterprise Applications: A Survey. *IEEE Trans. Ind. Inform.* **2014**, *10*, 35–42. https://doi.org/10.1109/tii.2012.2189221.

38. Mamdiwar, S.D.; Akshith, R.; Shakruwala, Z.; Chadha, U.; Srinivasan, K.; Chang, C.Y. Recent advances on IOT-Assisted Wearable Sensor Systems for healthcare monitoring. *Biosensors* **2021**, *11*, 372. https://doi.org/10.3390/bios11100372.

39. Patel, W.D.; Pandya, S.; Koyuncu, B.; Ramani, B.; Bhaskar, S.; Ghayvat, H. NXTGeUH: LoRaWAN based NEXT generation ubiquitous healthcare system for vital signs Monitoring falls detection. In Proceedings of the 2018 IEEE Punecon, Pune, India, 30 November–2 December 2018; pp. 1–8.

40. Ramya, C.M.; Shanmugaraj, M.; Prabakaran, R. Study on ZigBee technology. In Proceedings of the 3rd International Conference on Electronics Computer Technology, Kanyakumari, India, 8–10 April 2011; https://doi.org/10.1109/icectech.2011.5942102.

41. Purnama, B.; Sharipuddin; Kurniabudi; Budiarto, R.; Stiawan, D.; Hanapi, D. Monitoring Connectivity of Internet of Things Device on Zigbee Protocol. In Proceedings of the International Conference on Electrical Engineering and Computer Science, ICECOS 2018, Pangkal, Indonesia, 2–4 October 2018; pp. 351–356. https://doi.org/10.1109/icecos.2018.8605225.

42. Wang, S.; Li, L.; Wang, K.; Jones, J. E-business system integration: A systems perspective. *Inf. Technol. Manag.* **2012**, *13*, 233–249.

43. Xu, L. Enterprise Systems: State-of-the-art and future trends. *IEEE Trans. Ind. Informat.* **2011**, *7*, 630–640.

44. Lin, J.; Yu, W.; Zhang, N.; Yang, X.; Zhang, H.; Zhao, W. A Survey on Internet of Things: Architecture, Enabling Technologies, Security and Privacy, and Applications. *IEEE Internet Things J.* **2017**, *4*, 1125–1142. https://doi.org/10.1109/jiot.2017.2683200.

45. Hausenblas, M. Smart Phones and the Internet of Things. Available online: https://www.mapr.com/blog/smart-phones-and-internet-things (accessed on 3 August 2021).

46. Khaddar, M.A.E. Smartphone: The Ultimate IoT and IoE Device. IntechOpen. Available online: https://www.intechopen.com/chapters/56113 (accessed on 3 August 2021).

47. Simplifying IoT: Connecting, Commissioning, and Controlling with Near Field Communication (NFC). NFC Forum. Available online: http://nfc-forum.org/wp-content/uploads/2016/06/NFC_Forum_IoT_White_Paper_-v05.pdf (accessed on 10 March 2020).

48. Majumder, A.; Ghosh, S.; Goswami, J.; Bhattacharyya, B.K. NFC in IoT-Based Payment Architecture. *Internet Things* **2017**, 203–220. https://doi:10.1201/9781315156026-12.

49. Ramasamy, L.K.; Kadry, S. Internet of Things (IOT). *Blockchain Ind. Internet Things* **2021**. https://doi.org/10.1088/978-0-7503-3663-5ch1.

50. Laghari, A.A.; Wu, K.; Laghari, R.A.; Ali, M.; Khan, A.A. A review and state of art of internet of things (IOT). *Arch. Comput. Methods Eng.* **2021**, 1–19. https://doi.org/10.1007/s11831-021-09622-6.

51. Daniel, M.; Occhiogrosso, B. Practical aspects for the integration of 5G networks and IoT applications in smart cities environments. *Wirel. Commun. Mob. Comput.* **2019**, *2019*, 5710834.

52. Gregory, G. Bringing dark data into the light: Illuminating existing IoT data lost within your organization. *Bus. Horiz.* **2020**, *63*, 519–530.

53. Quek, T. The Advantages and Disadvantages of Internet of Things (IoT). Available online: https://www.linkedin.com/pulse/advantages-disadvantages-internet-things-iot-tommy-quek (accessed on 10 March 2020).

54. Caesarendra, W.; Pappachan, B.; Wijaya, T.; Lee, D.; Tjahjowidodo, T.; Then, D.; Manyar, O. An AWS Machine Learning-Based Indirect Monitoring Method for Deburring in Aerospace Industries towards Industry 4.0. *Appl. Sci.* **2018**, *8*, 2165. https://doi.org/10.3390/app8112165.

55. Antão, L.; Pinto, R.; Reis, J.P.; Gonçalves, G.M.; Naji, H.I.; Amer, W.; Maula, B.; Parkash, B.; Pai, A.; Tian, W.; et al. Fig. 1. Five Layer IOT Architecture. ResearchGate. 2018. Available onlone: https://www.researchgate.net/figure/Five-Layer-IoT-Architecture_fig1_324797771 (accessed on 19 December 2021).

56. Muccini, H.; Moghaddam, M.T. IOT architectural styles. *Softw. Archit.* **2018**, *11048*, 68–85. https://doi.org/10.1007/978-3-030-00761-4_5.

57. Mondal, D. The Internet of Thing (IoT) and Industrial Automation: A future perspective. *World J. Model. Simul.* **2019**, *15*, 140–149.

58. Lombardi, M.; Pascale, F.; Santaniello, D. Internet of things: A general overview between architectures, protocols and applications. *Information* **2021**, *12*, 87. https://doi.org/10.3390/info12020087.

59. Abdmeziem, M.R.; Tandjaoui, D.; Romdhani, I. Architecting the Internet of Things: State of the art. *Sens. Clouds* **2015**, *36*, 55–75.

60. Al-Fuqaha, A.I.; Guizani, M.; Mohammadi, M.; Aledhari, M.; Ayyash, M. Internet of Things: A survey on enabling technologies, protocols, and applications. *IEEE Commun. Surv. Tutor.* **2015**, *17*, 2347–2376.

61. Colaković, A.; Hadžialić, M. Internet of Things (IoT): A review of enabling technologies, challenges, and open research issues. *Comput. Netw.* **2018**, *144*, 17–39.

62. Guinard, D.; Trifa, V.; Karnouskos, S.; Spiess, P.; Savio, D. Interacting with the SOA-based internet of things: Discovery, query, selection, and on-demand provisioning of web services. *IEEE Trans. Serv. Comput.* **2010**, *3*, 223–235. https://doi.org/10.1109/tsc.2010.3.

63. Gama, K.; Touseau, L.; Donsez, D. Combining heterogeneous service technologies for building an internet of things middleware. *Comput. Commun.* **2012**, *35*, 405–417. https://doi.org/10.1016/j.comcom.2011.11.003.

64. Vermesan, O.; Eisenhauer, M.; Serrano, M.; Guillemin, P.; Sundmaeker, H.; Tragos, E.; Valiño, J.; Copigneaux, B.; Presser, M.; Aagaard, A.; et al. The Next Generation Internet of Things—Hyperconnectivity and Embedded Intelligence at the Edge. Available online: https://european-iot-pilots.eu/wp-content/uploads/2020/06/SRIA-2018_The_Next_Generation_IoT_Hyperconnectivity_and_Embedded_Intelligence_at_the_Edge_Research_Trends_IERC_2018_Cluster_eBook_978-87-7022-007-1_P_Web.pdf (accessed on 29 December 2021).

65. Waldner, J.B. *Nano-Informatique et Intelligence Ambiante*; Hermes Science Publications: Paris, France, 2006; ISBN 10:2746215160.

66. Qadir, Q.M.; Rashid, T.A.; Al-Salihi, N.K.; Ismael, B.; Kist, A.A.; Zhang, Z. Low Power Wide Area Networks: A Survey of Enabling Technologies, Applications and Interoperability Needs. *IEEE Access* **2018**, *6*, 77454–77473. https://doi.org/10.1109/access.2018.2883151.

67. Lampropoulos, G.; Siakas, K.; Anastasiadis, T. Internet of Things in the Context of Industry 4.0: An Overview. *Int. J. Entrep. Knowl.* **2019**, *7*, 4–19. https://doi.org/10.2478/ijek-2019-0001.

68. Akpakwu, G.A.; Silva, B.J.; Hancke, G.P.; Abu-Mahfouz, A.M. A Survey on 5G Networks for the Internet of Things: Communication Technologies and Challenges. *IEEE Access* **2018**, *6*, 3619–3647. https://doi.org/10.1109/access.2017.2779844.

69. Lin, J.; Yu, W.; Yang, X.; Yang, Q.; Fu, X.; Zhao, W. A Novel Dynamic En-Route Decision Real-Time Route Guidance Scheme in Intelligent Transportation Systems. In Proceedings of the 2015 IEEE 35th International Conference on Distributed Computing Systems ICDCS 2015, Columbus, OH, USA, 29 June–2 July 2015; pp. 61–72.

70. Hazarika, M.S.; Chowdhury, B.D. About Us | Xantra. Available online: http://xantra.in/about.html (accessed on 1 March 2020).

71. Zhang, L. Applications of the Internet of Things in the Medical Industry (Part 1): Digital Hospitals. 24 June 2018. Available online: https://dzone.com/articles/applications-of-the-internet-of-things-in-the-medi-1 (accessed on 22 February 2020).

72. Matthews, K. How IoT Is Enabling the Telemedicine of Tomorrow. 28 November 2018. Available online: https://www.iot-forall.com/how-iot-enables-tomorrows-telemedicine/ (accessed on 16 March 2020).

73. Vilamovska, A.M.; Hatziandreou, E.; Schindler, R.H.; Nassau, C.O.; Vries, H.; Krapelse, J. Study on the requirements and options for RFID application in healthcare Identifying areas for Radio Frequency Identification deployment in health care delivery: A review of relevant literature. Available online: https://www.rand.org/pubs/technical_reports/TR608.html (accessed on 20 August 2020).

74. Whitmore, A.; Agarwal, A.; Da Xu, L. The Internet of Things—A survey of topics and trends. *Inf. Syst. Front.* **2014**, *17*, 261–274. https://doi.org/10.1007/s10796-014-9489-2.

75. Karjagi, R.; Jindal, M. IoT in Healthcare Industry|IoT Applications in Healthcare—Wipro. Available online: https://www.wipro.com/business-process/what-can-iot-do-for-healthcare-/ (accessed on 15 March 2020).

76. Horwitz, L. Patient Health Data Is Increasingly Democratized–Despite Data Quality Issues. March 2020. Available online: https://www.iotworldtoday.com/2020/03/03/democratization-of-patient-health-data-empowers-despite-data-quality-issues/ (accessed on 30 March 2020).

77. Meola, A. IoT Healthcare in 2020: Companies, Devices, Use Cases and Market Stats. Available online: https://www.businessinsider.com/iot-healthcare?international=true&r=US&IR=T (accessed on 30 March 2020).

78. Work Apple Watch Series 5. Available online: https://www.apple.com/lae/apple-watch-series-5/workout/ (accessed on 11 September 2021).

79. Guo, B.; Yu, Z.; Zhou, X.; Zhang, D. Opportunistic IoT: Exploring the social side of the internet of things. In Proceedings of the IEEE 16th International Conference on Computer Supported Cooperative Work in Design, CSCWD 2012, Wuhan, China, 23–25 May 2012; pp. 925–929. https://doi.org/10.1109/cscwd.2012.6221932.

80. Mukhopadhyay, D.; Gupta, M.; Attar, T.; Chavan, P.; Patel, V. An attempt to develop an IOT based vehicle security system. In Proceedings of the 2018 IEEE International Symposium on Smart Electronic Systems (ISES) (Formerly iNiS), Hyderabad, India, 17–19 December 2018. https://doi.org/10.1109/ises.2018.00050.

81. WHO. *World Health Organization Report about Air Pollution WHO*; WHO: Geneva, Switzerland, 2016. Available online: https://apps.who.int/iris/bitstream/handle/10665/250141/9789241511353-eng.pdf (accessed on 28 June 2019).

82. Kaivonen, S.; Ngai, E.C.-H. Real-time air pollution monitoring with sensors on City Bus. *Digit. Commun. Netw.* **2020**, *6*, 23–30. https://doi.org/10.1016/j.dcan.2019.03.003.

83. Manna, S.; Bhunia, S.S.; Mukherjee, N. Vehicular pollution monitoring using IOT. In Proceedings of the International Conference on Recent Advances and Innovations in Engineering, ICRAIE 2014, Jaipur, India, 9–11 May 2014. https://doi.org/10.1109/icraie.2014.6909157.

84. Gubbi, J.; Buyya, R.; Marusic, S.; Palaniswami, M. Internet of Things (IoT): A vision, architectural elements, and future directions. *Future Gener. Comput. Syst.* **2013**, *29*, 1645–1660. https://doi.org/10.1016/j.future.2013.01.010.

85. Qadri, Y.A.; Nauman, A.; Zikria, Y.B.; Vasilakos, A.V.; Kim, S.W. The future of healthcare internet of things: A survey of emerging technologies. *IEEE Commun. Surv. Tutor.* **2020**, *22*, 1121–1167. https://doi.org/10.1109/comst.2020.2973314.

86. Gupta, D.; Bhatt, S.; Gupta, M.; Tosun, A.S. Future smart connected communities to fight COVID-19 outbreak. *Internet Things* **2021**, *13*, 100342. https://doi.org/10.1016/j.iot.2020.100342.

87. Technology Networks. Using the Internet of Things to Fight Virus Outbreaks. Available online: https://www.technologynetworks.com/immunology/articles/using-the-internet-of-things-to-fight-virus-outbreaks-331992 (accessed on 21 March 2020).

88. Perlstein, D. How Is IoT Helping to Fight the Coronavirus?—Axonize IoT and Coronavirus. Available online: https://www.axonize.com/blog/iot-automation/how-is-iot-helping-to-fight-the-coronavirus/?utm_source=trendemon&utm_medium=content&utm_campaign=flow (accessed on 28 March 2020).

89. Das, C.K.D.; Alam, M.W.; Hoque, M.I. A wireless heartbeat and Temperature Monitoring System for Remote Patients. In Proceedings of the International Conference on Mechanical Engineering and Renewable Energy, ICMERE 2014, Chittagong, Bangladesh, 1–3 May 2014. https://doi.org/10.13140/2.1.2568.9124.

90. Nall, R. What Is a Normally Temperature Range? Health News—Medical News Today. Available online: https://www.medicalnewstoday.com/articles/323819 (accessed on 20 April 2020).

91. Bellany, D.Y. Thermometer Guns' on Coronavirus Front Lines are Notoriously Not Accurate. The New York Times. Available online: https://www.nytimes.com/2020/02/14/business/coronavirus-temperature-sensor-guns.html (accessed on 26 November 2021).

92. Yousif, M.; Hewage, C.; Nawaf, L. IOT technologies during and beyond COVID-19: A comprehensive review. *Future Internet* **2021**, *13*, 105. https://doi.org/10.3390/fi13050105.

93. Bocetta, S. Hands-Free Everything? The Coronavirus Impact on IoT. Available online: https://www.globalsign.com/en/blog/hands-free-everything-coronavirus-impact-iot (accessed on 23 March 2021).

94. Western Shelter.COVID-19 Screening, Monitoring, and Isolation Part 1—Western Shelter. Available online: https://westernshelter.com/blog/2020/8/3/covid-19-screening-monitoring-and-isolation (accessed on 26 February 2021).

95. Koetsier, J. This Smart Home Gym Is the Future of Fitness. Available online: https://www.forbes.com/sites/johnkoetsier/2020/10/13/this-smart-home-gym-is-the-future-of-fitness/?sh=68691ae414bd (accessed on 26 February 2021).

96. Personal Use: Meet ADAMM—Health Care Originals. Available online: http://www.healthcareoriginals.com/personal (accessed on 2 March 2021).

97. Muthuramalingam, S.; Bharathi, A.; Kumar, S.R.; Gayathri, N.; Sathiyaraj, R.; Balamurugan, B. Iot based intelligent transportation system (iot-its) for global perspective: A case study. In *Intelligent Systems Reference Library*; Springer Science and Business Media LLC: Berlin/Heidelberg, Germany, 2019; Volume 154.

98. Chen, S.-W.; Gu, X.-W.; Wang, J.-J.; Zhu, H.-S. AIoT Used for COVID-19 Pandemic Prevention and Control. *Contrast Media Mol. Imaging* **2021**, *2021*, 3257035. https://doi.org/10.1155/2021/3257035.

99. Nasajpour, M.; Pouriyeh, S.; Parizi, R.M.; Dorodchi, M.; Valero, M.; Arabnia, H.R. Internet of things for current COVID-19 and future pandemics: An exploratory study. *J. Healthc. Inform. Res.* **2020**, *4*, 325–364. https://doi.org/10.1007/s41666-020-00080-6.

100. Hayes, A. Wearable Technology. Available online: https://www.investopedia.com/terms/w/wearable-technology.asp (accessed on 29 December 2021).

101. Smart Wearables Market to Generate $53bn Hardware Revenues by 2019. Available online: https://www.juniperresearch.com/press/smart-wearables-market-to-generate-53bn-hardware (accessed on 29 December 2021).

102. Berglund, M.E.; Duvall, J.; Dunne, L.E. A survey of the historical scope and current trends of wearable technology applications. In Proceedings of the 2016 ACM International Symposium on Wearable Computers, Heidelberg, Germany, 12–16 September 2016; pp. 40–43. https://doi.org/10.1145/2971763.2971796.

103. Fyntanidou, B.; Zouka, M.; Apostolopoulou, A.; Bamidis, P.D. IoT-based smart triage of COVID-19 suspicious cases in the Emergency Department. In Proceedings of the IEEE Globecom Workshops, Taipei, Taiwan, 7–11 December 2020; pp. 1–6.

104. Al Bassam, N.; Hussain, S.A.; Al Qaraghuli, A.; Khan, J.; Sumesh, E.P.; Lavanya, V. IoT based wearable device to monitor the signs of quarantined remote patients of COVID-19. *Inform. Med. Unlocked.* **2021**, *24*, 100588. https://doi.org/10.1016/j.imu.2021.100588.

105. Isabella, A.; Lekshmi, K.S.; Thamizhvaani, E.P.; Vishali, S. IOT Based Emergency Medical Services. *Int. J. Eng. Tech.* **2018**, *4*, 199–202.

106. Magurano, D. Development of a Scalable Architecture for Industrial Internet of Things. Available online: https://pdfs.semanticscholar.org/9ff6/7ba53c0fbaee90d5d27e43d5958b0ff4872f.pdf (accessed on 30 August 2020).

107. Miorandi, D.; Sicari, S.; De Pellegrini, F.; Chlamtac, I. Internet of things: Vision, applications and research challenges. *Ad Hoc Netw.* **2012**, *10*, 1497–1516. https://doi.org/10.1016/j.adhoc.2012.02.016.

108. Wang, S.; Wan, J.; Li, D.; Zhang, C. Implementing Smart Factory of Industrie 4.0: An Outlook. *Int. J. Distrib. Sens. Netw.* **2016**, *12*, 3159805. https://doi.org/10.1155/2016/3159805.

109. Lin, K.; Wang, W.; Bi, Y.; Qiu, M.; Hassan, M.M. Human localization based on inertial sensors and fingerprints in the Industrial Internet of Things. *Comput. Netw.* **2016**, *101*, 113–126. https://doi.org/10.1016/j.comnet.2015.11.012.

110. Jing, Q.; Vasilakos, A.V.; Wan, J.; Lu, J.; Qiu, D. Security of the Internet of Things: Perspectives and challenges. *Wirel. Netw.* **2014**, *20*, 2481–2501. https://doi.org/10.1007/s11276-014-0761-7.

111. Blanchet, M.; Rinn, T.; Thaden, G.; Thieulloy, G. *Industry 4.0. The New Industrial Revolution. How Europe Will Succeed. Hg V Roland Berg. Strategy Consult*; GmbH Münch. Abgerufen Am 1105 2014 Unter: Munich, Germany 2014.

112. Wollschlaeger, M.; Sauter, T.; Jasperneite, J. The Future of Industrial Communication: Automation Networks in the Era of the Internet of Things and Industry 4.0. *IEEE Ind. Electron. Mag.* **2017**, *11*, 17–27. https://doi.org/10.1109/mie.2017.2649104.

113. Mahmoodpour, M.; Lobov, A.; Lanz, M.; Makela, P.; Rundas, N. Role-based visualization of industrial IoT-based systems. In Proceedings of the 14th IEEE/ASME International Conference on Mechatronic and Embedded Systems and Applications, MESA 2018, Oulu, Finland, 2–4 July 2018. https://doi.org/10.1109/mesa.2018.8449183.

114. SlideModel. Industry 4.0 PowerPoint Template. Available online: https://slidemodel.com/templates/industry-4-0-powerpoint-template/ (accessed on 16 March 2020).

115. Lee, J.; Bagheri, B.; Kao, H.A. A Cyber-Physical Systems architecture for Industry 4.0-based manufacturing systems. *Manuf. Lett.* **2015**, *3*, 18–23. https://doi.org/10.1016/j.mfglet.2014.12.001.

116. Lasi, H.; Fettke, P.; Kemper, H.G.; Feld, T.; Hoffmann, M. Industry 4.0. *Bus. Inf. Syst. Eng.* **2014**, *4*, 239–242. Available online: https://link.springer.com/article/10.1007/s12599-014-0334-4#auth-1 (accessed on 1 May 2020).

117. Zhong, R.Y.; Xu, X.; Klotz, E.; Newman, S.T. Intelligent Manufacturing in the Context of Industry 4.0: A Review. *Engineering* **2017**, *3*, 616–630. https://doi.org/10.1016/j.eng.2017.05.015.

118. Batista, N.C.; Melício, R.; Mendes, V.M.F. Services enabler architecture for smart grid and smart living services providers under industry 4.0. *Energy Build.* **2017**, *141*, 16–27. https://doi.org/10.1016/j.enbuild.2017.02.039.

119. Zhao, F.; Sun, Z.; Jin, H. Topic-centric and semantic-aware retrieval system for internet of things. *Inf. Fusion* **2015**, *23*, 33–42. https://doi.org/10.1016/j.inffus.2014.01.001.

120. Liu, J.; Yan, Z.; Yang, L.T. Fusion–An aide to data mining in Internet of Things. *Inf. Fusion* **2015**, *23*, 1–2. https://doi.org/10.1016/j.inffus.2014.08.001.

121. Kalsoom, T.; Ahmed, S.; Rafi-ul-Shan, P.M.; Azmat, M.; Akhtar, P.; Pervez, Z.; Imran, M.A.; Ur-Rehman, M. Impact of IOT on Manufacturing Industry 4.0: A new triangular systematic review. *Sustainability* **2021**, *13*, 12506. https://doi.org/10.3390/su132212506.

122. Sanchez, L.M.; Nagi, R. A review of agile manufacturing systems. *Int. J. Prod. Res.* **2001**, *39*, 3561–3600.

123. Dallasega, P.; Rauch, E.; Linder, C. Industry 4.0 as an enabler of proximity for construction supply chains: A systematic literature review. *Comput. Ind.* **2018**, *99*, 205–225.

124. Kamble, S.S.; Gunasekaran, A.; Sharma, R. Analysis of the driving and dependence power of barriers to adopt Industry 4.0 in Indian manufacturing industry. *Comput. Ind.* **2018**, *101*, 107–119.

125. Gottge, S.; Menzel, T.; Forslund, H. Industry 4.0 technologies in the purchasing process. *Ind. Manag. Data Syst.* **2020**, *120*, 730–748.

126. Javaid, S.; Sufian, A.; Pervaiz, S.; Tanveer, M. Smart Traffic Management System using internet of things. In Proceedings of the 20th International Conference on Advanced Communication Technology, ICACT 2018, Chuncheon, Korea, 11–14 February 2018; https://doi.org/10.23919/icact.2018.8323769.

127. Sumia, L.; Ranga, V. Intelligent traffic management system for prioritizing emergency vehicles in a smart city. *Int. J. Eng.* **2018**, *31*, 278–283.

128. Pyykonen, P.; Laitinen, J.; Viitanen, J.; Eloranta, P.; Korhonen, T. IOT for intelligent traffic system. In Proceedings of the IEEE 9th International Conference on Intelligent Computer Communication and Processing, ICCP 2013, Cluj-Napoca, Romania, 5–7 September 2013; https://doi.org/10.1109/iccp.2013.6646104.

129. Afzal, B.; Umair, M.; Asadullah Shah, G.; Ahmed, E. Enabling IOT platforms for social IOT applications: Vision, feature mapping, and challenges. *Future Gener. Comput. Syst.* **2019**, *92*, 718–731. https://doi.org/10.1016/j.future.2017.12.002.

130. Nasr, E.; Kfoury, E.; Khoury, D. An IOT approach to vehicle accident detection, reporting, and Navigation. In Proceedings of the IEEE International Multidisciplinary Conference on Engineering Technology, IMCET 2016, Beirut, Lebanon, 2–4 November 2016. https://doi.org/10.1109/imcet.2016.7777457.

131. Raj, J.T.; Sankar, J. IOT based Smart School Bus Monitoring and Notification System. In Proceedings of the IEEE Region 10 Humanitarian Technology Conference, R10-HTC 2017, Dhaka, Bangladesh, 21–23 December 2017. https://doi.org/10.1109/r10-htc.2017.8288913.

132. Drath, R.; Horch, A. Industrie 4.0: Hit or Hype? [Industry Forum]. *IEEE Ind. Electron. Mag.* **2014**, *8*, 56–58. https://doi.org/10.1109/mie.2014.2312079.

133. Sreenivasulu, R.; Chalamalasetti, S.R. Applicability of Industrial Internet of Things (IIoT) in Lean Manufacturing: A brief Study. *Prod. Sched.* **2019**, *10*, 22–26. Available online: https://www.researchgate.net/publication/337286599_Applicability_of_Industrial_Internet_of_Things_IIoT_in_Lean_Manufacturing_A_brief_Study (accessed on 16 December 2021).

134. Harrell, C. The Internet of Things and Control System Architecture. Available online: http://blog.aac.advantech.com/th e-internet-of-things-and-control-system-architecture/ (accessed on 1 July 2020).

135. Capello, F.; Toja, M.; Trapani, N. A Real-Time Monitoring Service based on Industrial Internet of Things to manage agrifood logistics. In Proceedings of the 6th International Conference on Information Systems, Logistics and Supply Chain ILS Conference, Bordeaux, France, 1–4 June 2016.

136. Hiter, S. Industrial Internet of Things (IIoT) Market Size & Forecast. 2022. Available online: https://www.datamation.com/trends/industrial-internet-of-things-iiot-market/ (accessed on 16 December 2021).

137. Evans, D. The Internet of Things How the Next Evolution of the Internet Is Changing Everything. *Cisco Internet Bus. Solut. Group (IBSG)* **2011**, *1*, 1–11. Available online: https://www.cisco.com/c/dam/en_us/about/ac79/docs/innov/IoT_IBSG_0411FINAL.pdf (accessed on 10 February 2021).

138. Afshar, V. HuffPost Is Now a Part of Verizon Media. Available online: https://www.huffpost.com/entry/cisco-enterprises-are-leading-the-internet-of-things_b_59a41fcee4b0a62d0987b0c6 (accessed on 14 June 2020).

139. Statista Research Department. IoT: Number of Connected Devices Worldwide 2012–2025. Available online: https://www.statista.com/statistics/471264/iot-number-of-connected-devices-worldwide/ (accessed on 10 February 2020).

140. Patil, P.; Tawade, P.; Samudre, S.; Mali, S.; Pardeshi, S. Survey on Internet of Things (IoT): Tools and Technologies. *Int. J. Sci. Res. Comput. Sci. Eng. Inf. Technol. (IJSRCSEIT)* **2018**, *3*, 1108–1112. Available online: http://ijsrcseit.com/CSEIT1833352 (accessed on 3 April 2021).

141. Heutger, M. A Shared Journey: Customer-Centric Innovation-Huawei Case Studies. Huawei Enterprise. Available online: https://e.huawei.com/en/case-studies/global/2017/201709070922 (accessed on 16 December 2021).

142. Cisco. Solutions-Cisco IOT Helps Nissan Transform Car Production Operations. Cisco. Available online: https://www.cisco.com/c/en/us/solutions/collateral/internet-of-things/nissan-motors-case-study.html (accessed on 16 December 2021).

143. Gartner Gartner Says 5.8 Billion Enterprise and Automotive IoT Endpoints Will Be in Use in 2020. Available online: https://www.gartner.com/en/newsroom/press-releases/2019-08-29-gartner-says-5-8-billion-enterprise-and-automotive-io (accessed on 3 April 2020).

144. Presser, M.; Gluhak, A. The Internet of things: Connecting the real world with the digital world. Eurescom Message—The Magazine for Telecom Insiders. 2009. Volume 2. Available online: http://www.eurescom.eu/message (accessed on 1 January 2021).

145. Ray, P.P. A survey on Internet of Things architectures. *J. King Saud Univ.-Comput. Inf. Sci.* **2018**, *30*, 291–319. https://doi.org/10.1016/j.jksuci.2016.10.003.

146. Frost & Sullivan Best Practices Award. Radio-Frequency Identification for Retail Self Checkouts—Europe. 2019; pp. 1–19. Available online: https://www.nordicid.com/wp-content/uploads/Frost-Sullivan-Nordic-ID-New-Product-Innovation-2019-report.pdf (accessed on 10 December 2021).

147. Industrial Internet of Things Market Size Report. 2021–2028. Available online: https://www.grandviewresearch.com/industry-analysis/industrial-internet-of-things-iiot-market (accessed on 16 December 2021).

148. Shen, W.; Norrie, D.H. Agent-Based Systems for Intelligent Manufacturing: A State-of-the-Art Survey. *Knowl. Inf. Syst.* **1999**, *1*, 129–156. https://doi.org/10.1007/bf03325096.

149. Li, X.; Li, D.; Wan, J.; Vasilakos, A.V.; Lai, C.-F.; Wang, S. A review of industrial wireless networks in the context of Industry 4.0. *Wirel. Netw.* **2015**, *23*, 23–41. https://doi.org/10.1007/s11276-015-1133-7.

150. Lund, D.; Carrie, C.; Turner, V.; Morales, M. Worldwide and Regional Internet of Things (IoT) 2014–2020 Forecast: A Virtuous Circle of Proven Value and Demand. IDC Analyze the Future, 1–27. Available online: http://branden.biz/wp-content/uploads/2017/06/IoT-worldwide_regional_2014-2020-forecast.pdf (accessed on 10 January 2020).

151. Krishnamachari, B.; Power, J.; Shahabi, C.; Kim, S.H. *IoT Marketplace: A Data and API Market for IoT Devices [PDF]*; University of Southern California, Los Angeles, CA, USA, 2017.

152. Saputhanthri, A.; Alwis, C.D.; Liyanage, M. Emergence of Blockchain Based IOT Marketplaces. Available online: https://www.researchgate.net/profile/Madhusanka-Liyanage/publication/350835689_Emergence_of_Blockchain_based_IoT_Marketplaces/links/6080a4c8907dcf667bb5ae4f/Emergence-of-Blockchain-based-IoT-Marketplaces.pdf (accessed on 19 December 2021).